\documentclass[12pt]{article}

\usepackage{amsmath}
\usepackage{amssymb}
\usepackage{graphicx}
\usepackage{cite}
\usepackage{bm}
\usepackage{color}

\makeatletter

\numberwithin{equation}{section}

\allowdisplaybreaks

\newcounter{aff}

\makeatother

\begin{document}
\begin{titlepage}
\begin{flushright}
{\footnotesize YITP-18-115, OCU-PHYS 489, KIAS-P18098}
\end{flushright}
\begin{center} 
{\LARGE\bf
Symmetry Breaking in Quantum Curves\\[6pt]
and Super Chern-Simons Matrix Models}\\
\bigskip\bigskip
{\large Naotaka Kubo\,\footnote{\tt naotaka.kubo@yukawa.kyoto-u.ac.jp},
\quad
Sanefumi Moriyama\,\footnote{\tt moriyama@sci.osaka-cu.ac.jp},
\quad
Tomoki Nosaka\,\footnote{\tt nosaka@yukawa.kyoto-u.ac.jp}}\\
\bigskip
${}^{*}$\,{\it Center for Gravitational Physics, Yukawa Institute for Theoretical Physics,}\\
{\it Kyoto University, Sakyo-ku, Kyoto 606-8502, Japan}\\[3pt]
${}^{\dagger}$\,{\it Department of Physics, Graduate School of Science, Osaka City University,}\\
${}^\dagger$\,{\it Nambu Yoichiro Institute of Theoretical and Experimental Physics (NITEP),}\\
${}^\dagger$\,{\it Osaka City University Advanced Mathematical Institute (OCAMI),}\\
{\it Sumiyoshi-ku, Osaka 558-8585, Japan}\\[3pt]
${}^{\ddagger}$\,{\it School of Physics, Korea Institute for Advanced Study,}\\
{\it Dongdaemun-gu, Seoul 02455, Korea}
\end{center}
\begin{abstract}
It was known that quantum curves and super Chern-Simons matrix models correspond to each other.
From the viewpoint of symmetry, the algebraic curve of genus one, called the del Pezzo curve, enjoys symmetry of the exceptional algebra, while the super Chern-Simons matrix model is described by the free energy of topological strings on the del Pezzo background with the symmetry broken.
We study the symmetry breaking of the quantum cousin of the algebraic curve and reproduce the results in the super Chern-Simons matrix model.
\end{abstract}
\centering\includegraphics[scale=0.4,angle=90]{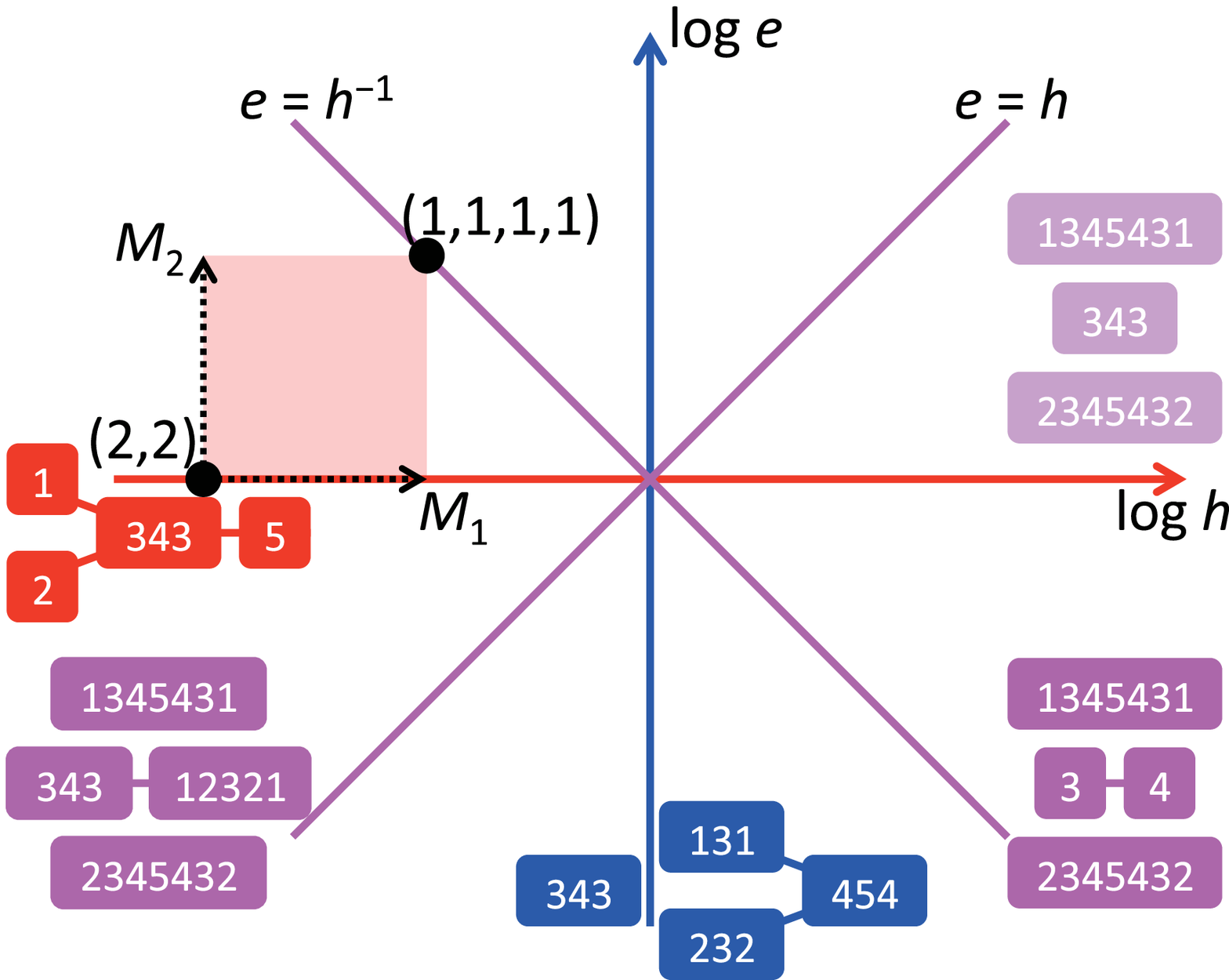}
\end{titlepage}

\setcounter{footnote}{0}

\tableofcontents

\section{Introduction}

In the semi-classical analysis of quantum physics, curves appear as the phase space orbit leading to the semi-classical Bohr-Sommerfeld quantization condition.
From the viewpoint of full quantum physics, apparently these curves are technical tools and need to be quantized eventually.
Since in algebraic geometry curves are defined by zeros of polynomial rings, in quantization by replacing the polynomial rings by quantum operators, the study of zeros switches smoothly to spectral problems of the quantum operators.

Recently the study of quantization of curves attracts renewed attention due to the important role it plays in understanding the integrability of gauge theories \cite{ADKMV,NS,MiMo,ACDKV}.
The interplay between curves and gauge theories continues to the three-dimensional M2-brane physics \cite{ABJM}.
In \cite{MP} it was found that the grand canonical partition function of the M2-branes on the background ${\mathbb C}^4/{\mathbb Z}_k$ can be rewritten as a spectral determinant of a quantum-mechanical operator associated with the geometry ${\mathbb{P}}^{1}\times{\mathbb{P}}^{1}$, following preceding works \cite{MPtop,DMP1,HKPT,DMP2,FHM}.
After further studies of the spectral determinant in \cite{KEK,HMO1,PY,HMO2,CM,HMO3} finally it was conjectured \cite{HMMO} that the grand potential of the M2-branes is expressed as the free energy of the topological string theory on the local ${\mathbb{P}}^{1}\times{\mathbb{P}}^{1}$ geometry.
The rank deformations with the inclusion of fractional M2-branes \cite{HLLLP2,ABJ} were studied in \cite{AHS,H,MM,HO} and found to match the conjectured topological string free energy.

The computation was further generalized to many superconformal Chern-Simons matrix models describing the worldvolume theory of the M2-branes on other orbifold \cite{HM,MN1,MN2,MN3,HHO,MNN,MNY} or orientifold \cite{MS1,Hosp,Oosp,MS2,MN5} backgrounds and to many spectral determinants associated with other curves \cite{GHM1,OZ}.
On the matrix model side, we can compute the models with or without rank deformations with similar techniques and find that the results fit in the conjecture with different choices of K\"ahler parameters and BPS indices on the background geometry, although it is difficult to explain these geometrical data.
On the curve side, the general structure of the correspondence is much clearer, though, besides the difficulty in the interpretation in terms of the M2-branes, it was also difficult to compute directly the kernels of the spectral operators for general parameters of the curves until recently with the important progress in \cite{KaMa,KMZ}.
All of these difficulties on the both sides prevent us from studying the correspondence clearly.

To overcome the difficulties, the viewpoint of symmetry is crucial.
On the curve side, among others, the special class of curves of ultimate interest and importance are those of genus one called del Pezzo curves, which are known to enjoy the symmetries of the exceptional algebra.
On the matrix model side, by studying rank deformations of matrix models corresponding to curves of genus one, in several cases, the K\"ahler parameters and the BPS indices were identified \cite{MNN} and these geometrical data were further interpreted from the symmetry breaking in \cite{MNY}.
Although the symmetry breaking patterns were identified, the explanation of them was missing.

In this paper, we promote the discussions for the classical del Pezzo curves to quantum curves.
We define the quantum curves for our setup and study how the symmetry is realized and how the symmetry breaking happens.
We find that the breaking patterns are completely consistent with the previous results in \cite{MNY} from the superconformal Chern-Simons matrix models.

The organization of this paper is as follows.
We first review the analysis of the superconformal Chern-Simons matrix models in the next section to explain our motivation.
After that, in section \ref{sec_QC}, we define the quantum curve and study its quantum symmetry using an example of the $D_5$ curve.
In section \ref{sec_SB} we identify the superconformal Chern-Simons matrix models in the quantum curve and study its symmetry breaking.
We turn to a different example of the $E_7$ curve in section \ref{sec_SEC}.
Finally we conclude with some discussions on future directions.
Appendix \ref{e7weyl} is devoted to technical details on the construction of the Weyl group.

\section{Superconformal Chern-Simons matrix models}

In this section, we review the superconformal Chern-Simons matrix models.
The main purpose of this section is to explain our motivation of studying quantum curves.

It was proposed \cite{ABJM,HLLLP2,ABJ} that the ${\cal N}=6$ superconformal Chern-Simons theory with gauge group U$(N_{1})_{k}\times$U$(N_{2})_{-k}$ (with the subscripts $k,-k$ denoting the Chern-Simons levels) and two pairs of bifundamental matters describes the worldvolume theory of $\min(N_{1},N_{2})$ M2-branes and $|N_{2}-N_{1}|$ fractional M2-branes on the target space ${\mathbb{C}}^{4}/{\mathbb{Z}}_{k}$.
With the localization techniques \cite{KWY}, the partition function, as well as the one-point functions (and hopefully the two-point functions \cite{KuMo}) of the half-BPS Wilson loop in the ${\cal N}=6$ superconformal Chern-Simons theory on $S^{3}$, which is originally defined with the infinite-dimensional path integral, reduces to a finite-dimensional matrix integration.

There are many generalizations for this matrix model.
For example, by regarding the quiver diagram of the ABJM theory as the Dynkin diagram of the affine Lie algebra $\widehat{A}_{1}$, there are other generalized theories with quiver diagrams of affine simply-laced Lie algebras and it is known that they also preserve the ${\cal N}=3$ superconformal symmetries \cite{GAH}.\footnote{See \cite{ADF,MN4} progress in the study of the matrix models for the $\widehat{D}_{r}$ quiver diagram.}
Especially, it was found \cite{IK4} that, for the $\widehat{A}_{r}$ quiver diagram with the gauge group U$(N)^{r+1}$, as long as the levels $k_{a}$ for $a=1,2,\cdots,r+1$ are given by 
\begin{align}
k_{a}=\frac{k}{2}(s_{a}-s_{a-1}),\quad s_{a}=\pm 1,
\label{sN4}
\end{align}
(with the cyclic identification $s_{0}=s_{r+1}$), the superconformal Chern-Simons theory enjoys the supersymmetry enhancement to ${\cal N}=4$.
Hence, the ${\cal N}=4$ theories can be characterized by recording $\{s_{a}\}_{a=1}^{r+1}$ with the order.
Following the same localization technique, the partition functions of these theories are given clearly by associating the vector multiplets (or vertices in the quiver diagram) and the hypermultiplets (or edges in the quiver diagram) respectively with
\begin{align}
\prod_{m<m'}^{N}\biggl(2\sinh\frac{\lambda_{a,(m)}-\lambda_{a,(m')}}{2}\biggr)^{2},\quad\prod_{m,n=1}^{N}\biggl(2\cosh\frac{\lambda_{a,(m)}-\lambda_{a+1,(n)}}{2}\biggr)^{-2},
\end{align}
and integrating all of the variables $\lambda_{a,(m)}$ with
\begin{align}
D\lambda_{a,(m)}=\frac{d\lambda_{a,(m)}}{2\pi}e^{\frac{ik_{a}}{4\pi}\sum_{m=1}^{N}\lambda_{a,(m)}^{2}}.
\end{align}
In \cite{MN1} the model specified by the $\pm1$ alignment
\begin{align}
\{s_{a}\}_{a=1}^{r+1}&=\bigl(\overbrace{+1,\cdots,+1}^{q_{1}},
\overbrace{-1,\cdots,-1}^{p_{1}},
\overbrace{+1,\cdots,+1}^{q_{2}},
\overbrace{-1,\cdots,-1}^{p_{2}},\cdots
\bigr),
\end{align}
was named the $(q_{1},p_{1},q_{2},p_{2},\cdots
)$ model
and the grand canonical partition function of the model $\Xi_{k}(z)=\Xi_{k}^{(q_{1},p_{1},q_{2},p_{2},\cdots)}(z)$ defined from the partition function $Z_{k}(N)=Z_{k}^{(q_{1},p_{1},q_{2},p_{2},\cdots)}(N)$ by 
\begin{align}
\Xi_{k}(z)=\sum_{N=0}^{\infty}z^{N}Z_{k}(N),
\end{align}
was found to be given by 
\begin{align}
\Xi_{k}(z)=\det\bigl(1+z{\widehat H}^{-1}\bigr),
\label{fredholm}
\end{align}
with $\widehat{H}=\widehat{H}^{(q_{1},p_{1},q_{2},p_{2},\cdots)}$
of the model given by
\begin{align}
\widehat{H}
=\biggl(2\cosh\frac{\widehat{q}}{2}\biggr)^{q_{1}}
\biggl(2\cosh\frac{\widehat{p}}{2}\biggr)^{p_{1}}
\biggl(2\cosh\frac{\widehat{q}}{2}\biggr)^{q_{2}}
\biggl(2\cosh\frac{\widehat{q}}{2}\biggr)^{p_{2}}\cdots.
\label{hamiltonian}
\end{align}
Here $\widehat{q}$ and $\widehat{p}$ are canonical operators satisfying the commutation relation $\left[\widehat{q},\widehat{p}\right]=2\pi ik$.

In a series of works \cite{MN1,MN2,MN3,HHO,MNN,MNY} following the study of the ABJM matrix model \cite{HMO2,CM,HMO3,HMMO} it was further found that the reduced grand potential $J_{k}(\mu)=J_{k}^{(q_{1},p_{1},q_{2},p_{2},\cdots)}(\mu)$ of a class of the models (of genus one) defined by
\begin{align}
\sum_{n=-\infty}^{\infty}e^{J_{k}(\mu+2\pi in)}=\Xi_{k}(e^{\mu}),
\end{align}
is split into the perturbative part, the worldsheet instanton part and membrane instanton part, $J_{k}(\mu)=J_{k}^{\text{pert}}(\mu)+J^{\text{WS}}(\mu)+J^{\text{MB}}(\mu)$ and if we redefine the chemical potential $\mu$ into $\mu_{\text{eff}}$ \cite{HMO3} and further into the K\"ahler parameters ${\bm{T}}$, the instanton parts are described by the free energy of topological strings 
\begin{align}
J_{k}^{\text{WS}}(\mu) & =\sum_{j_{\text{L}},j_{\text{R}}}\sum_{\bm{d}}N_{j_{\text{L}},j_{\text{R}}}^{\bm{d}}\sum_{n=1}^{\infty}\frac{s_{\text{R}}\sin2\pi g_{\text{s}}ns_{\text{L}}}{n(2\sin\pi g_{\text{s}}n)^{2}\sin2\pi g_{\text{s}}n}e^{-n{\bm{d}}\cdot{\bm{T}}},\nonumber\\
J_{k}^{\text{MB}}(\mu) & =\sum_{j_{\text{L}},j_{\text{R}}}\sum_{\bm{d}}N_{j_{\text{L}},j_{\text{R}}}^{\bm{d}}\sum_{n=1}^{\infty}\frac{\partial}{\partial g_{\text{s}}}\biggl[g_{\text{s}}\frac{-\sin\frac{\pi n}{g_{\text{s}}}s_{\text{L}}\sin\frac{\pi n}{g_{\text{s}}}s_{\text{R}}}{4\pi n^{2}(\sin\frac{\pi n}{g_{\text{s}}})^{3}}e^{-n\frac{{\bm{d}}\cdot{\bm{T}}}{g_{\text{s}}}}\biggr],
\end{align}
(with $s_{\text{L}/\text{R}}=2j_{\text{L}/\text{R}}+1$) on a target space which can be read off from \eqref{hamiltonian}.

Especially, it turns out that the target spaces for the $(1,1)$, $(2,2)$, $(1,1,1,1)$, $(2,1)$ and $(2,1,2,1)$ models are the del Pezzo curves of genus one, known to be classified by the exceptional Lie algebra $E_{n}$.
From a careful analysis of the exact values of the partition function in \cite{MN1,MN3,MNN}, it was found \cite{MNY} that the $(2,2)$, $(1,1,1,1)$ and $(2,1)$ models correspond to the $E_{5}=D_{5}$ del Pezzo curve at the moduli where the Weyl symmetry of the $D_{5}$ algebra is broken respectively to those of the subalgebras\footnote{As we explain later, the remaining symmetry $(A_{1})^{4}$ for the $(1,1,1,1)$ model identified in \cite{MNY} should be corrected by $(A_{1})^{3}$.} $D_{4}$, $(A_{1})^{3}$ and $A_{3}$, while the $(2,1,2,1)$ model corresponds to the $E_{7}$ del Pezzo curve at the modulus where the $E_{7}$ algebra is broken to the subalgebra $D_{5}\times A_{1}$.
Namely, for example for the $(2,2)$ model and the $(1,1,1,1)$ model, since the total BPS indices $N_{j_{\text{L}},j_{\text{R}}}^{d}=\sum_{|\bm{d}|=d}N_{j_{\text{L}},j_{\text{R}}}^{\bm{d}}$ at each degree $d$ were computed in \cite{HKP}, our task reduces to identifying the K\"ahler parameters ${\bm{T}}$ and the split of the total BPS indices at each degree $d$.
This was performed in \cite{MNN} and it was further found in \cite{MNY} that, by regarding the BPS indices as representations of the $D_{5}$ algebra, the introduction of the K\"ahler parameters amounts to identifying ``the Higgs fields acquiring expectation values'' and the split of the total BPS indices corresponds to the decomposition of the representations of the $D_{5}$ algebra into those of the unbroken subalgebras.

\begin{table}[!t]
\begin{centering}
\begin{tabular}{|c|c|c|c|}
\hline
$d$ & $(j_{\text{L}},j_{\text{R}})$ & BPS & $(-1)^{d-1}\sum_{d_{\text{I}}}\bigl(\sum_{d_{\text{II}}}N_{j_{\text{L}},j_{\text{R}}}^{(d,d_{\text{I}},d_{\text{II}})}\bigr)_{d_{\text{I}}}$\\
\hline
\hline
$1$ & $(0,0)$ & $16$ & $8_{+1}+8_{-1}$\\
\hline
$2$ & $(0,\frac{1}{2})$ & $10$ & $1_{+2}+8_{0}+1_{-2}$\\
\hline
$3$ & $(0,1)$ & $16$ & $8_{+1}+8_{-1}$\\
\hline
$4$ & $(0,\frac{1}{2})$ & $1$ & $1_{0}$\\
\cline{2-4}
 & $(0,\frac{3}{2})$ & $45$ & $8_{+2}+29_{0}+8_{-2}$\\
\cline{2-4}
 & $(\frac{1}{2},2)$ & $1$ & $1_{0}$\\
\hline
\end{tabular}
\par\end{centering}
\caption{The split of the BPS indices on the $D_{5}$ del Pezzo curve for the $(2,2)$ model.
The split is interpreted as the decomposition of the $D_{5}$ representations into the $D_{4}$ subalgebra, as in ${\bf 16}\to({\bf 8_{s/c}})_{+1}+({\bf 8_{s/c}})_{-1}$, ${\bf 10}\to({\bf 1})_{+2}+({\bf 8_{v}})_{0}+({\bf 1})_{-2}$ and ${\bf 45}\to({\bf 8_{v}})_{+2}+({\bf 28})_{0}+({\bf 1})_{0}+({\bf 8_{v}})_{-2}$.}
\label{d4}
\end{table}

More concretely, the rank deformations of the $(2,2)$ model and the $(1,1,1,1)$ model, which are connected by the Hanany-Witten effect, were studied intensively in \cite{MNN}.
For the $(2,2)$ model with the rank deformations U$(N)_{k}\times$U$(N+M_{\text{I}})_{0}\times$U$(N+2M_{\text{I}})_{-k}\times$U$(N+M_{\text{I}})_{0}$, the K\"ahler parameters and the string coupling constant $g_{\text{s}}$ in the instanton exponents $e^{-{\bm{d}}\cdot{\bm{T}}}$ and $e^{-{\bm{d}}\cdot{\bm{T}}/g_{\text{s}}}$ are 
\begin{align}
T^{\pm}=\frac{\mu_{\text{eff}}}{k}\pm\pi i\biggl(1-\frac{M_{\text{I}}}{k}\biggr),\quad
g_s=\frac{1}{k},
\end{align}
and the BPS indices forming the representations of the $D_{5}$ algebra are broken to representations of $D_{4}$ (see table \ref{d4} for the split of the BPS indices).
Furthermore, for the $(2,2)$ model with the rank deformations 
\begin{equation}
\mathrm{U}(N+M_{\text{II}})_{k}\times\mathrm{U}(N+M_{\text{I}})_{0}\times\mathrm{U}(N+2M_{\text{I}}+M_{\text{II}})_{-k}\times\mathrm{U}(N+M_{\text{I}})_{0},\label{rankDef}
\end{equation}
which is connected to the $(1,1,1,1)$ model without rank deformations at $(M_{\text{I}},M_{\text{II}})=(k/2,k/2)$ through the Hanany-Witten effect, the K\"ahler parameters are
\begin{align}
T_{1}^{\pm}&=\frac{\mu_{\text{eff}}}{k}\pm\pi i\biggl(1-\frac{M_{\text{I}}}{k}-\frac{2M_{\text{II}}}{k}\biggr),\nonumber\\
T_{2}^{\pm}&=\frac{\mu_{\text{eff}}}{k}\pm\pi i\biggl(1-\frac{M_{\text{I}}}{k}\biggr),\nonumber\\
T_{3}^{\pm}&=\frac{\mu_{\text{eff}}}{k}\pm\pi i\biggl(1-\frac{M_{\text{I}}}{k}+\frac{2M_{\text{II}}}{k}\biggr), \label{kahlerDeform}
\end{align}
and the BPS indices in the representations of the $D_{4}$ algebra are further split into representations of the $(A_{1})^{3}$ algebra (see table \ref{a1^3} for the further split of the BPS indices).
Hence, from table \ref{d4} it was found that the symmetry for the $(2,2)$ model without rank deformations is broken to $D_{4}$ while from table \ref{a1^3} the symmetry for the $(1,1,1,1)$ model is further broken to $(A_{1})^{3}$.

Even though the symmetry is broken to $(A_1)^3$ for general rank deformations of the $(2,2)$ model and the $(1,1,1,1)$ model, we can see an accidental symmetry enhancement for the $(1,1,1,1)$ model without rank deformations.
Since the $(1,1,1,1)$ model without rank deformations corresponds to $(M_{\text{I}},M_{\text{II}})=(k/2,k/2)$, the instanton exponent is given by 
\begin{align}
{\bm{d}}\cdot{\bm{T}}
=d\frac{\mu_{\text{eff}}}{k}
+\pi i\biggl(d_{\text{I}}\biggl(1-\frac{M_{\text{I}}}{k}\biggr)
-d_{\text{II}}\frac{2M_{\text{II}}}{k}\biggr)
=d\frac{\mu_{\text{eff}}}{k}+\widetilde{d}\frac{\pi i}{2},
\label{kalher1111}
\end{align}
where we have defined the total u$(1)$ degree and the two Cartan u$(1)$ charges which break the symmetries as
\begin{align}
d=\sum_{i=1}^{3}(d_{i}^{+}+d_{i}^{-}),\quad
d_{\text{I}}=(d_{1}^{+}+d_{2}^{+}+d_{3}^{+})-(d_{1}^{-}+d_{2}^{-}+d_{3}^{-}),\quad
d_{\text{II}}=(d_{1}^{+}-d_{1}^{-})-(d_{3}^{+}-d_{3}^{-}),
\end{align}
as well as the special combination of the u$(1)$ charges $\widetilde{d}$ characterizing the $(1,1,1,1)$ model without rank deformations and the unbroken u$(1)$ charge $\overline{d}$ as\footnote{
The combination of the u$(1)$ charges $\overline{d}=2(d_1^+-d_1^-)+(d_2^+-d_2^-)$ exchanges among the four K\"ahler parameters $T_1^\pm=\frac{\mu_\text{eff}}{k}\mp\frac{\pi i}{2}$ and $T_2^\pm=\frac{\mu_\text{eff}}{k}\pm\frac{\pi i}{2}$, while leaving the remaining two $T_3^\pm=\frac{\mu_\text{eff}}{k}\pm\frac{3\pi i}{2}$ fixed.
} 
\begin{align}
\widetilde{d}=d_{\text{I}}-2d_{\text{II}},\quad
\overline{d}=d_{\text{I}}+d_{\text{II}}.
\end{align}

Then, the accidental symmetry enhancement is observed as follows.
The adjoint representation of the $D_{5}$ algebra decomposes as 
\begin{align}
{\bf 45} & \to({\bf 8_{v}})_{+2}+({\bf 28})_{0}+({\bf 1})_{0}+({\bf 8_{v}})_{-2},
\end{align}
in the breaking $D_{5}\to(D_{4})_{d_{\text{I}}}$ and further decompositions of various $D_{4}$ representations into $(A_{1})^{4}$ are given by 
\begin{align}
{\bf 28} & \to({\bf 3},{\bf 1},{\bf 1},{\bf 1})+({\bf 1},{\bf 3},{\bf 1},{\bf 1})+({\bf 1},{\bf 1},{\bf 3},{\bf 1})+({\bf 1},{\bf 1},{\bf 1},{\bf 3})+({\bf 2},{\bf 2},{\bf 2},{\bf 2}),\nonumber\\
{\bf 8_{v}} & \to({\bf 2},{\bf 2},{\bf 1},{\bf 1})+({\bf 1},{\bf 1},{\bf 2},{\bf 2}).
\end{align}
The last factor of $A_{1}$ is broken and the u(1) charge is denoted by $d_{\text{II}}$.
After expressing the two u(1) charges $d_{\text{I}}$ and $d_{\text{II}}$ in terms of the charges $\overline{d}$ and $\widetilde{d}$, each representation of the last factor of $(A_{1})_{\overline{d}}$ in the unbroken symmetry $(A_{1})^{3}$ combines into the representations of $A_{2}$ as
\begin{align}
{\bf 8}\to{\bf 2}_{+3}+{\bf 3}_{0}+{\bf 1}_{0}+{\bf 2}_{-3},\quad{\bf 3}\to{\bf 1}_{+2}+{\bf 2}_{-1},\quad\overline{{\bf 3}}\to{\bf 2}_{+1}+{\bf 1}_{-2},\quad{\bf 1}\to{\bf 1}_{0}
\end{align}
in $A_{2}\to(A_{1})_{\overline{d}}$.
Finally the decomposition of the $D_{5}$ adjoint representation into $(A_{1}\times A_{1}\times A_{2})_{\widetilde d}$ is given by
\begin{align}
{\bf 45}\to&({\bf 1},{\bf 1},\overline{{\bf 3}})_{+4}+({\bf 2},{\bf 2},{\bf 3})_{+2}+({\bf 3},{\bf 1},{\bf 1})_{0}+({\bf 1},{\bf 3},{\bf 1})_{0}+({\bf 1},{\bf 1},{\bf 8})_{0}+({\bf 1},{\bf 1},{\bf 1})_{0}\nonumber\\
& +({\bf 2},{\bf 2},\overline{{\bf 3}})_{-2}+({\bf 1},{\bf 1},{\bf 3})_{-4},
\label{A2enhance}
\end{align}
which implies that the symmetry $(A_{1})^{3}$ is further enhanced to $(A_{1})^{2}\times A_{2}=A_{1}\times A_{1}\times A_{2}$ in the $(1,1,1,1)$ model without rank deformations.

\begin{table}[!t]
\begin{centering}
\begin{tabular}{|c|c|c|c|c|}
\hline 
$d$ & $(j_{\text{L}},j_{\text{R}})$ & $d_{\text{I}}$ & BPS & $(-1)^{d-1}\sum_{d_{\text{II}}}\bigl(N_{j_{\text{L}},j_{\text{R}}}^{(d,d_{\text{I}},d_{\text{II}})}\bigr)_{d_{\text{II}}}$\\
\hline 
\hline 
$1$ & $(0,0)$ & $\pm1$ & $8$ & $2_{+1}+4_{0}+2_{-1}$\\
\hline 
$2$ & $(0,\frac{1}{2})$ & $0$ & $8$ & $2_{+1}+4_{0}+2_{-1}$\\
\cline{3-5}
 &  & $\pm2$ & $1$ & $1_{0}$\\
\hline 
$3$ & $(0,1)$ & $\pm1$ & $8$ & $2_{+1}+4_{0}+2_{-1}$\\
\hline 
$4$ & $(0,\frac{1}{2})$ & $0$ & $1$ & $1_{0}$\\
\cline{2-5} 
 & $(0,\frac{3}{2})$ & $0$ & $29$ & $1_{+2}+8_{+1}+11_{0}+8_{-1}+1_{-2}$\\
\cline{3-5}
 &  & $\pm2$ & $8$ & $2_{+1}+4_{0}+2_{-1}$\\
\cline{2-5}
 & $(\frac{1}{2},2)$ & $0$ & $1$ & $1_{0}$\\
\hline 
\end{tabular}
\par\end{centering}
\caption{The split of the BPS indices on the $D_{5}$ del Pezzo curve for the $(1,1,1,1)$ model.
The split is interpreted as the decomposition of the $D_{4}$ representations into the $(A_{1})^{4}$ subalgebra, as in
${\bf 8_{v}}\to({\bf 2},{\bf 2},{\bf 1},{\bf 1})+({\bf 1},{\bf 1},{\bf 2},{\bf 2})$,
${\bf 8_{s}}\to({\bf 2},{\bf 1},{\bf 2},{\bf 1})+({\bf 1},{\bf 2},{\bf 1},{\bf 2})$,
${\bf 8_{c}}\to({\bf 2},{\bf 1},{\bf 1},{\bf 2})+({\bf 1},{\bf 2},{\bf 2},{\bf 1})$
and
${\bf 28}\to({\bf 3},{\bf 1},{\bf 1},{\bf 1})+({\bf 1},{\bf 3},{\bf 1},{\bf 1})+({\bf 1},{\bf 1},{\bf 3},{\bf 1})+({\bf 1},{\bf 1},{\bf 1},{\bf 3})+({\bf 2},{\bf 2},{\bf 2},{\bf 2})$,
where the last $A_{1}$ factor contributes as the u$(1)$ charge in the subscript (and hence is broken).}
\label{a1^3} 
\end{table}

\section{Quantum curve}\label{sec_QC}

In this section we define carefully what we mean by quantum curves and study the typical example of the $D_{5}$ del Pezzo curve.

We define a quantum algebraic curve to be the spectral problem of a polynomial quantum operator $\widehat{H}$ generated by $\widehat{Q}=e^{\widehat{q}}$ and $\widehat{P}=e^{\widehat{p}}$ where $\widehat{q}$ and $\widehat{p}$ are the canonical operators of coordinates and momenta satisfying the canonical commutation relation $[\widehat{q},\widehat{p}]=i\hbar$.
Since the similarity transformation, the adjoint action by $\widehat{G}$,
\begin{align}
\widehat{H}\sim\widehat{G}\widehat{H}\widehat{G}^{-1},
\label{similarity}
\end{align}
typically does not affect the spectral problem, we define the quantum algebraic curve with the identification of all the similarity transformations.
As in the classical algebraic curve, the curve is studied within a linear combination of a certain class of the independent operators $\widehat{Q}^{m}\widehat{P}^{n}$ with $(m,n)\in\mathbb{Z}^2$.
Note that in the quantization, the order of the operators is important and we adopt the normal ordering such that $\widehat{Q}$ is in the left and $\widehat{P}$ is in the right.
The set of $(m,n)$ with non-vanishing coefficients is often referred to as the Newton polygon.

\begin{figure}[!t]
\centering\includegraphics[scale=0.6,angle=-90]{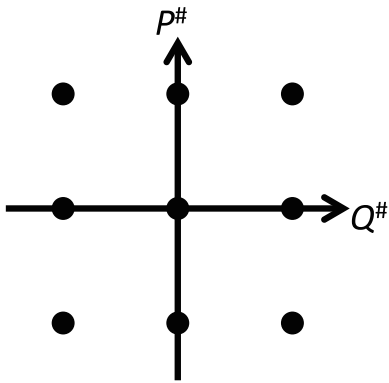}
\caption{The Newton polygon of the $D_{5}$ del Pezzo curve.}
\label{d5np} 
\end{figure}

\begin{figure}[!t]
\centering\includegraphics[scale=0.6,angle=90]{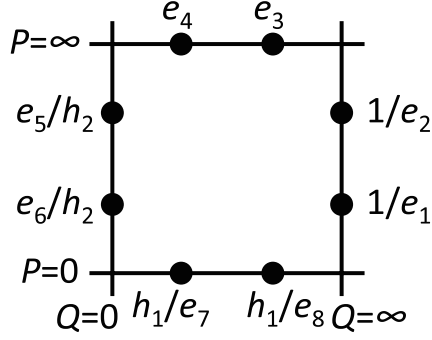}
\caption{The asymptotic values of the $D_{5}$ del Pezzo curve, $(\infty,e_{1}^{-1})$,
$(\infty,e_{2}^{-1})$, $(e_{3},\infty)$, $(e_{4},\infty)$, $(0,h_{2}^{-1}e_{5})$,
$(0,h_{2}^{-1}e_{6})$, $(h_{1}e_{7}^{-1},0)$, $(h_{1}e_{8}^{-1},0)$.
The four lines denote ``lines at infinity'' $\widehat{Q}=\infty$,
$\widehat{P}=\infty$, $\widehat{Q}=0$ and $\widehat{P}=0$ respectively.}
\label{d5asymp} 
\end{figure}

The classical algebraic curve of genus one is called the del Pezzo curve and is known to be classified by the exceptional algebra $E_{n}$.
As one of simple and abundant cases, here we mainly consider the quantization of the $E_{5}=D_{5}$ curve, where the quantum curve or the quantum Hamiltonian is a linear combination of the independent operators
\begin{align}
\widehat{Q}^{m}\widehat{P}^{n},\quad m=-1,0,1,\quad n=-1,0,1,
\end{align}
(see figure \ref{d5np} for the Newton polygon of the $D_{5}$ del Pezzo curve).
Instead of fixing the coefficients, as in the classical case \cite{KNY} (see sections 8.2.5 and 8.4.4), it is often convenient to fix the asymptotic values of the curve
\begin{align}
(\infty,e_{1}^{-1}),(\infty,e_{2}^{-1}),(e_{3},\infty),(e_{4},\infty),(0,h_{2}^{-1}e_{5}),(0,h_{2}^{-1}e_{6}),(h_{1}e_{7}^{-1},0),(h_{1}e_{8}^{-1},0),
\end{align}
(see figure \ref{d5asymp} for the asymptotic values).
Each two points out of the eight points are the solutions to the quadratic equations obtained by setting $\widehat{Q}\to\infty$, $\widehat{P}\to\infty$, $\widehat{Q}\to0$ and $\widehat{P}\to0$ respectively.
In other words, the eight values are the asymptotic values of the dual graph of the Newton polygon.
Due to the Vieta's formulas on products of roots, the eight points are not independent and should be subject to the constraint
\begin{align}
\prod_{i=1}^{8}e_{i}=h_{1}^{2}h_{2}^{2}.\label{constraint}
\end{align}
Then, our quantum curve is given by
\begin{align}
\widehat{H}/\alpha=\begin{array}{ccc}
\widehat{Q}\widehat{P} & -(e_{3}+e_{4})\widehat{P} & +e_{3}e_{4}\widehat{Q}^{-1}\widehat{P}\\
-(e_{1}^{-1}+e_{2}^{-1})\widehat{Q} & +E/\alpha & -h_{2}^{-1}e_{3}e_{4}(e_{5}+e_{6})\widehat{Q}^{-1}\\
+(e_{1}e_{2})^{-1}\widehat{Q}\widehat{P}^{-1} & -h_{1}(e_{1}e_{2})^{-1}(e_{7}^{-1}+e_{8}^{-1})\widehat{P}^{-1} & +h_{1}^{2}(e_{1}e_{2}e_{7}e_{8})^{-1}\widehat{Q}^{-1}\widehat{P}^{-1},
\end{array}
\label{curve}
\end{align}
where the coefficient of the last term $\widehat{Q}^{-1}\widehat{P}^{-1}$ can be alternatively expressed as $h_{1}^{2}(e_{1}e_{2}e_{7}e_{8})^{-1}=h_{2}^{-2}e_{3}e_{4}e_{5}e_{6}$ due to \eqref{constraint}.
Note that the classical algebraic curve is defined from the zeros of the curve and characterized by their asymptotic zeros.
For the quantum case, the asymptotic zeros are obtained only after the normal ordering.

This curve enjoys a lot of symmetries.
Especially our labelling of the curve is redundant and the same curve can be realized by different
choices of the parameters.
For example, we use ten parameters $h_{1},h_{2}$ and $e_{1},\cdots,e_{8}$ to describe the eight asymptotic values and apparently two degrees of freedom can be fixed arbitrarily.
Also, by the similarity transformation \eqref{similarity} generated by $\widehat{G}=e^{\frac{ia}{\hbar}\widehat{p}}$ or $\widehat{G}=e^{-\frac{ib}{\hbar}\widehat{q}}$, it is clear that a curve and the same curve with ${\widehat Q},{\widehat P}$ rescaled as
\begin{align}
(\widehat{Q},\widehat{P})\to(A\widehat{Q},\widehat{P}),\quad
(\widehat{Q},\widehat{P})\to(\widehat{Q},B\widehat{P}),
\label{rescaling}
\end{align}
(with $A=e^{a}$ and $B=e^{b}$) should be identified.
Using these two rescalings we can further fix two degrees of freedom.
After the identification, aside from the parameters $\alpha$ and $E$, we have ten parameters subject to four continuous symmetries and one constraint \eqref{constraint}, which leaves only five parameters.

After identifying these continuous gauge symmetries, there also remain discrete gauge symmetries, which should be clarified.
The analysis for the classical algebraic curve is well-known and explained carefully for example in \cite{KNY}.
Here we study the same problem for the quantum curve.

\begin{figure}[!t]
\centering\includegraphics[scale=0.6,angle=-90]{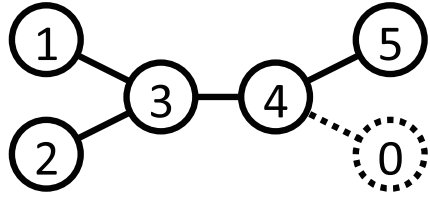}
\caption{The Dynkin diagram of the $D_{5}$ algebra.}
\label{d5dd}
\end{figure}

Classically the $D_{5}$ del Pezzo curve enjoys the Weyl symmetry of $D_{5}$, which is basically generated by $s_{1},s_{2},s_{3},s_{4},s_{5}$ exchanging the asymptotical points \cite{KNY}
\begin{align}
s_{1}:h_{1}e_{7}^{-1}\leftrightarrow h_{1}e_{8}^{-1},\quad
s_{2}:e_{3}\leftrightarrow e_{4},\quad
s_{3}:e_{3}\leftrightarrow h_{1}e_{7}^{-1},\quad
s_{4}:e_{1}^{-1}\leftrightarrow h_{2}^{-1}e_{5},\quad
s_{5}:e_{1}^{-1}\leftrightarrow e_{2}^{-1},
\end{align}
and has $2^{4}\times5!=1920$ elements in total
(see figure \ref{d5dd} for the Dynkin diagram of $D_5$ and the numbering of the roots).
Though in our setup the affine root does not appear, we can introduce the lowest root
\begin{align}
s_{0}:h_{2}^{-1}e_{5}\leftrightarrow h_{2}^{-1}e_{6},
\end{align}
to complete the affine Dynkin diagram.
Of course, this is not necessary because the lowest root is generated by the simple roots as $s_0=s_4s_3s_2s_5s_4s_3s_1s_3s_4s_5s_2s_3s_4$.

To provide the Weyl symmetry explicitly, in the following we adopt the gauge fixing condition
\begin{align}
e_{2}=e_{4}=e_{6}=e_{8}=1,
\end{align}
using the two degrees of freedom in the redundant description of the eight points with ten parameters and the other two degrees of freedom in the continuous rescaling as explained in \eqref{rescaling}.
Then, the constraint \eqref{constraint} becomes
\begin{align}
h_{1}^{2}h_{2}^{2}=e_{1}e_{3}e_{5}e_{7}.
\label{constraint2}
\end{align}
We often drop $e_7$ with
\begin{align}
e_{7}=h_{1}^{2}h_{2}^{2}(e_{1}e_{3}e_{5})^{-1},
\label{e7he}
\end{align}
to display the transformations unambiguously.

After the gauge fixing, it is not difficult to realize that the exchanges of $s_{1}$, $s_{2}$, $s_{5}$ and $s_{0}$ are given as
\begin{align}
& s_{1}:(h_{1},h_{2},e_{1},e_{3},e_{5},e_{7};\alpha)\mapsto(h_{1}e_{7}^{-1},h_{2},e_{1},e_{3},e_{5},e_{7}^{-1};\alpha),\nonumber\\
& s_{2}:(h_{1},h_{2},e_{1},e_{3},e_{5},e_{7};\alpha)\mapsto(h_{1}e_{3}^{-1},h_{2},e_{1},e_{3}^{-1},e_{5},e_{7};e_{3}\alpha),\nonumber\\
& s_{5}:(h_{1},h_{2},e_{1},e_{3},e_{5},e_{7};\alpha)\mapsto(h_{1},h_{2}e_{1}^{-1},e_{1}^{-1},e_{3},e_{5},e_{7};e_{1}^{-1}\alpha),\nonumber\\
& s_{0}:(h_{1},h_{2},e_{1},e_{3},e_{5},e_{7};\alpha)\mapsto(h_{1},h_{2}e_{5}^{-1},e_{1},e_{3},e_{5}^{-1},e_{7};\alpha),
\end{align}
as in the classical case. For $s_{3}$ and $s_{4}$ the situation is more complicated.
For $s_{3}$ we apply the canonical transformation,
\begin{align}
\widehat{Q}'=\widehat{Q},\quad
\widehat{P}'=(\widehat{Q}-e_{3})\widehat{P}(\widehat{Q}-h_{1}e_{7}^{-1})^{-1},
\end{align}
which can be obtained by applying the similarity transformation \eqref{similarity} generated by
\begin{align}
\widehat{G}=e^{F_{3}(\widehat{q})-F_{7}(\widehat{q})},
\end{align}
with $F_{3}(q)$ and $F_{7}(q)$ defined by
\begin{align}
e^{F_{3}(q)-F_{3}(q-i\hbar)}=e^{q}-e_{3},\quad e^{F_{7}(q+i\hbar)-F_{7}(q)}=e^{q}-h_{1}e_{7}^{-1}.
\end{align}
Indeed, by using the formula $e^{\mp {\widehat p}}f({\widehat q})e^{\pm {\widehat p}}=f({\widehat q}\pm i\hbar)$ repeatedly one can show
\begin{align}
&\widehat G\widehat P\widehat G^{-1}
=e^{F_3({\widehat q})-F_7({\widehat q})}e^{\widehat p}e^{-F_3({\widehat q})+F_7({\widehat q})}
\nonumber\\
&\quad=e^{F_3({\widehat q})-F_3({\widehat q}-i\hbar)}e^{\widehat p}e^{F_7({\widehat q})-F_7({\widehat q}+i\hbar)}
=(\widehat Q-e_3)\widehat P(\widehat Q-h_1e_7^{-1})^{-1}.
\end{align}
Then, after the normal ordering, we find that the terms in $\widehat{H}/\alpha$ proportional to $\widehat{P}$ and those proportional to $\widehat{P}^{-1}$ are respectively given by ($q=e^{i\hbar}$)
\begin{align}
\widehat{Q}^{-1}(\widehat{Q}-e_{3})(\widehat{Q}-1)\widehat{P} & =\widehat{Q}'^{-1}(\widehat{Q}'-qh_{1}e_{7}^{-1})(\widehat{Q}'-1)(q^{-1}\widehat{P}'),\nonumber\\
e_{1}^{-1}\widehat{Q}^{-1}(\widehat{Q}-h_{1}e_{7}^{-1})(\widehat{Q}-h_{1})\widehat{P}^{-1} & =e_{1}^{-1}\widehat{Q}'^{-1}(\widehat{Q}'-q^{-1}e_{3})(\widehat{Q}'-h_{1})(q^{-1}\widehat{P}')^{-1}.
\end{align}
Similarly for $s_{4}$ we apply the similarity transformation
\begin{align}
\widehat{Q}'=(\widehat{P}-h_{2}^{-1}e_{5})^{-1}\widehat{Q}(\widehat{P}-e_{1}^{-1}),\quad\widehat{P}'=\widehat{P},
\end{align}
and perform a similar normal ordering.
These transformations imply that
\begin{align}
s_{3}:(h_{1},h_{2},e_{1},e_{3},e_{5},e_{7};\alpha) & \mapsto
(h_{1},qh_{1}h_{2}(e_{3}e_{7})^{-1},e_{1},qh_{1}e_{7}^{-1},e_{5},qh_{1}e_{3}^{-1};\alpha),
\nonumber\\
s_{4}:(h_{1},h_{2},e_{1},e_{3},e_{5},e_{7};\alpha) & \mapsto
(h_{1}h_{2}(qe_{1}e_{5})^{-1},h_{2},h_{2}(qe_{5})^{-1},e_{3},h_{2}(qe_{1})^{-1},e_{7};\alpha).
\end{align}
Using the constraint \eqref{e7he}, we find
\begin{align}
s_{1}:(h_{1},h_{2},e_{1},e_{3},e_{5};\alpha) & \mapsto
\biggl(\frac{e_{1}e_{3}e_{5}}{h_{1}h_{2}^{2}},h_{2},e_{1},e_{3},e_{5};\alpha\biggr),\nonumber\\
s_{2}:(h_{1},h_{2},e_{1},e_{3},e_{5};\alpha) & \mapsto
\biggl(\frac{h_{1}}{e_{3}},h_{2},e_{1},\frac{1}{e_{3}},e_{5};e_{3}\alpha\biggr),\nonumber\\
s_{3}:(h_{1},h_{2},e_{1},e_{3},e_{5};\alpha) & \mapsto
\biggl(h_{1},\frac{qe_{1}e_{5}}{h_{1}h_{2}},e_{1},\frac{qe_{1}e_{3}e_{5}}{h_{1}h_{2}^{2}},e_{5};\alpha\biggr),\nonumber\\
s_{4}:(h_{1},h_{2},e_{1},e_{3},e_{5};\alpha) & \mapsto\biggl(\frac{h_{1}h_{2}}{qe_{1}e_{5}},h_{2},\frac{h_{2}}{qe_{5}},e_{3},\frac{h_{2}}{qe_{1}};\alpha\biggr),\nonumber\\
s_{5}:(h_{1},h_{2},e_{1},e_{3},e_{5};\alpha) & \mapsto\biggl(h_{1},\frac{h_{2}}{e_{1}},\frac{1}{e_{1}},e_{3},e_{5};\frac{\alpha}{e_{1}}\biggr),\nonumber\\
s_{0}:(h_{1},h_{2},e_{1},e_{3},e_{5};\alpha) & \mapsto\biggl(h_{1},\frac{h_2}{e_5},e_{1},e_{3},\frac{1}{e_5};\alpha\biggr).\label{s12345}
\end{align}
It is not difficult to see the algebraic relations
\begin{align}
& s_{1}^{2}=s_{2}^{2}=s_{3}^{2}=s_{4}^{2}=s_{5}^{2}=1,\nonumber\\
& (s_{1}s_{2})^{2}=(s_{1}s_{4})^{2}=(s_{1}s_{5})^{2}=(s_{2}s_{4})^{2}=(s_{2}s_{5})^{2}=(s_{3}s_{5})^{2}=1,\nonumber\\
& (s_{1}s_{3})^{3}=(s_{2}s_{3})^{3}=(s_{3}s_{4})^{3}=(s_{4}s_{5})^{3}=1.
\label{algrel}
\end{align}
By comparing with the general relations of the Weyl group $(s_is_j)^{r+2}=1$ for two different simple roots connected by $r$ edges, the relations \eqref{algrel} indicate that the transformations generate the Weyl group of $D_{5}$ in figure \ref{d5dd}.

Apparently, in the transformations \eqref{s12345} only $s_3$ and $s_4$ contain the quantum deformation parameter $q$ explicitly and, by setting $q=1$, the transformations reproduce those for the classical curves.
It is, then, natural to ask whether the transformations for the quantum curves essentially change from the classical ones.
To answer this question, let us redefine $h_1$ and $h_2$ by
\begin{align}
\overline{h}_{1}=qh_{1},\quad\overline{h}_{2}=q^{-1}h_{2}.
\label{hbar}
\end{align}
After the redefinition, the transformations $s_{3}$ and $s_{4}$ become
\begin{align}
s_{3}:(\overline{h}_{1},\overline{h}_{2},e_{1},e_{3},e_{5};\alpha) & \mapsto\biggl(\overline{h}_{1},\frac{e_{1}e_{5}}{\overline{h}_{1}\overline{h}_{2}},e_{1},\frac{e_{1}e_{3}e_{5}}{\overline{h}_{1}\overline{h}_{2}^{2}},e_{5};\alpha\biggr),\nonumber\\
s_{4}:(\overline{h}_{1},\overline{h}_{2},e_{1},e_{3},e_{5};\alpha) & \mapsto\biggl(\frac{\overline{h}_{1}\overline{h}_{2}}{e_{1}e_{5}},h_{2},\frac{\overline{h}_{2}}{e_{5}},e_{3},\frac{\overline{h}_{2}}{e_{1}};\alpha\biggr),
\label{s3s4}
\end{align}
and the other transformations are unaffected by the change of variables.
Hence we conclude that the only change from the transformations for the classical curves is the shift of the parameters \eqref{hbar}.

This fact implies that we can regard the parameter space of the curve as the root or weight space and identify these transformations as the standard Weyl actions, reflections by the simple root vectors $\alpha$,
\begin{align}
s_{\alpha}(v)=v-\alpha\frac{2(\alpha,v)}{(\alpha,\alpha)},
\label{weylrefl}
\end{align}
where $v$ is an element of a five-dimensional space and $(\cdot,\cdot)$ is a bilinear form in the space.
To identify the simple root vectors in the parameter space of the curve, we also prepare the fundamental weight vectors $\omega_i$ ($1\le i\le 5$), which are defined as the dual basis of the coroot vectors, $(\omega_i,\alpha^\vee_j)=\delta_{ij}$, with the coroot vectors being $\alpha^\vee_i=2\alpha_i/(\alpha_i,\alpha_i)$.
Then, we find that the root vectors are expanded by the fundamental weight vectors with the coefficients given by the Cartan matrix $A_{ij}=(\alpha_i,\alpha^\vee_j)$,
\begin{align}
\alpha_{i}=A_{ij}\omega_{j},
\label{alphaAomega}
\end{align}
and that transformation $s_{\alpha_i}$ acts on $\omega_j$ as
\begin{align}
s_{\alpha_{i}}(\omega_{j})=\omega_{j}-\delta_{ij}\alpha_{i},
\label{D5relations}
\end{align}
with no sum over $i$.
Now, it turns out that our task for finding simple roots and fundamental weights is to solve \eqref{alphaAomega}  and \eqref{D5relations} simultaneously under the identification $s_{\alpha_i}(v)=s_i(v)$ along with the explicit form of the Cartan matrix of $D_5$
\begin{align}
A=\begin{pmatrix}
2 & 0 & -1 & 0 & 0\\
0 & 2 & -1 & 0 & 0\\
-1 & -1 & 2 & -1 & 0\\
0 & 0 & -1 & 2 & -1\\
0 & 0 & 0 & -1 & 2
\end{pmatrix}.
\end{align}
Then, we find that the final results of the identification are given as
\begin{align}
&\alpha_{1}=(1,0,0,0,0),\quad\omega_1=(1,-1,0,0,-1),\nonumber\\
&\alpha_{2}=(1,0,0,2,0),\quad\omega_2=(1,-1,0,1,-1),\nonumber\\
&\alpha_{3}=(0,-1,0,-1,0),\quad\omega_3=(1,-2,0,0,-2),\nonumber\\
&\alpha_{4}=(-1,0,-1,0,-1),\quad\omega_4=(0,-1,0,0,-2),\nonumber\\
&\alpha_{5}=(0,1,2,0,0),\quad\omega_5=(0,0,1,0,-1),
\end{align}
where we have represented the parameters of curves by $(\log \overline{h}_1,\log \overline{h}_2,\log e_1,\log e_3,\log e_5)$. 
Note that in this expression, our symmetries of the algebraic curve $s_i$ \eqref{s12345}, \eqref{s3s4} reduce to the standard Weyl action $s_{\alpha_i}$ \eqref{weylrefl}.

\section{Symmetry breaking}\label{sec_SB}

After establishing the Weyl symmetries of the quantum $D_{5}$ del Pezzo curve, we start our study of the symmetry breaking.
For the $(2,2)$ model and the $(1,1,1,1)$ model with the expressions of the quantum operator
\begin{align}
\widehat{H}^{(2,2)} & =(\widehat{Q}^{\frac{1}{2}}+\widehat{Q}^{-\frac{1}{2}})^{2}(\widehat{P}^{\frac{1}{2}}+\widehat{P}^{-\frac{1}{2}})^{2},\nonumber\\
\widehat{H}^{(1,1,1,1)} & =(\widehat{Q}^{\frac{1}{2}}+\widehat{Q}^{-\frac{1}{2}})(\widehat{P}^{\frac{1}{2}}+\widehat{P}^{-\frac{1}{2}})(\widehat{Q}^{\frac{1}{2}}+\widehat{Q}^{-\frac{1}{2}})(\widehat{P}^{\frac{1}{2}}+\widehat{P}^{-\frac{1}{2}}),
\end{align}
after the shift $\widehat{q}\to\widehat{q}+\pi i$ and $\widehat{p}\to\widehat{p}-\pi i$, generated by the similarity transformations \eqref{rescaling},
we can easily identify the parameters
\begin{align}
(h_{1},h_{2},e_{1},e_{3},e_{5},e_{7};\alpha)^{(2,2)} & =(1,1,1,1,1,1;1),\nonumber\\
(h_{1},h_{2},e_{1},e_{3},e_{5},e_{7};\alpha)^{(1,1,1,1)} & =(1,1,q^{-\frac{1}{2}},q^{\frac{1}{2}},q^{-\frac{1}{2}},q^{\frac{1}{2}};q^{-\frac{1}{4}}).
\label{heparameter}
\end{align}
For the study of the $(2,1)$ model we need to rescale $\widehat{p}$ by $2$, 
\begin{align}
\widehat{H}^{(2,1)} & =(\widehat{Q}^{\frac{1}{2}}+\widehat{Q}^{-\frac{1}{2}})^2(\widehat{P}+\widehat{P}^{-1}).
\end{align}
Then, the parameters are
\begin{align}
(h_{1},h_{2},e_{1},e_{3},e_{5},e_{7};\alpha)^{(2,1)} & =(1,1,-1,1,-1,1;1).
\label{heparameter21}
\end{align}
Hence, the main question is, out of the $1920$ elements of the $D_{5}$ Weyl group, which elements leave these parameters including $\alpha$ invariant and what group these elements form.
We can easily generate the $1920$ elements with a computer by subsequently acting the transformations $s_{1},s_{2},s_{3},s_{4},s_{5}$ and choosing only the transformations which do not appear previously.
We can then act these $1920$ transformations on the parameters of the various models \eqref{heparameter}, \eqref{heparameter21} and pick up the invariant transformations.

\begin{figure}[!t]
\centering\includegraphics[scale=0.6,angle=90]{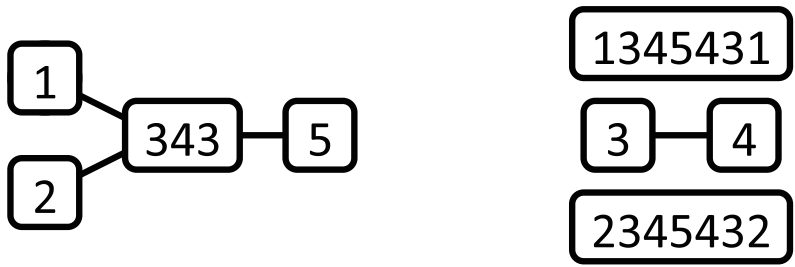}
\caption{The Dynkin diagram of the $D_4$ subalgebra within the original $D_5$ algebra which preserves the $(2,2)$ model without rank deformations (Left) and that of the $(A_1)^2\times A_2$ subalgebra which preserves the $(1,1,1,1)$ model without rank deformations (Right).}
\label{invalg}
\end{figure}

Let us search for the transformations leaving the parameters for these models invariant.
For the $(2,2)$ model, out of the $1920$ elements, we find that there are $192$ elements leaving the parameter invariant.
If we look closely, we further find that, among them, the four transformations
\begin{align}
s_{1},\quad s_{2},\quad s_{3}s_{4}s_{3},\quad s_{5},
\label{D4generators}
\end{align}
satisfy the relations
\begin{align}
(s_{1}s_{3}s_{4}s_{3})^{3}=(s_{2}s_{3}s_{4}s_{3})^{3}=(s_{5}s_{3}s_{4}s_{3})^{3}=1,
\end{align}
and generate $2^{3}\times4!=192$ different elements.
Hence we conclude that the invariant subgroup leaving the $(2,2)$ model is the $D_{4}$ Weyl group (see figure \ref{invalg} for the Dynkin diagram).
The result matches with the studies of the superconformal Chern-Simons matrix models as in table \ref{d4}.

We can further ask which subspace in the parameter space $(h_{1},h_{2},e_{1},e_{3},e_{5})$ enjoys the same $D_{4}$ symmetry as the $(2,2)$ model without rank deformations for arbitrary values of $\alpha$.
Due to the expression of the transformation $s_3s_4s_3$
\begin{align}
s_{3}s_{4}s_{3}:(h_{1},h_{2},e_{1},e_{3},e_{5};\alpha)\mapsto\biggl(\frac{1}{h_{2}},\frac{1}{h_{1}},\frac{e_{1}}{h_{1}h_{2}},\frac{e_{3}}{h_{1}h_{2}},\frac{e_{5}}{h_{1}h_{2}};\alpha\biggr),\label{s343}
\end{align}
the condition $h_{1}h_{2}=1$ is required.
Along with the actions of $s_{1}$, $s_{2}$ and $s_{5}$ in \eqref{s12345}, we further find the conditions $e_{1}=e_{3}=e_{5}=1$.
Namely, the subspace in the parameter space $({h}_{1},{h}_{2},e_{1},e_{3},e_{5},e_7)$ enjoying the same $D_{4}$ symmetry as the $(2,2)$ model without rank deformations is parametrized by
\begin{align}
\{(\overline{h}_{1},\overline{h}_{2},e_{1},e_{3},e_{5},e_7)=(h^{-1},h,1,1,1,1)\}.
\label{space1}
\end{align}
We find that, instead of the original parameter of $(h_1,h_2,e_1,e_3,e_5,e_7)$, in discussing the subspace with the symmetry enhancement, it is convenient to use the redefined parameter $(\overline{h}_1,\overline{h}_2,e_1,e_3,e_5,e_7)$ introduced in \eqref{hbar}.

Note that the symmetry breaking does not mean that the $D_{5}$ Weyl symmetry disappears completely.
Even though the broken symmetries do not leave the parameters invariant, since the transformations come from the similarity transformations, the new parameters share the same spectrum.
In the analogy of the spontaneous symmetry breaking, the broken symmetry is realized ``non-linearly''.
More concretely, with the broken symmetries, the vacuum expectation value is mapped to other equivalent values which share the same symmetry breaking.
In fact since the order of the $D_5$ Weyl group is $1920$ and the order of the invariant $D_4$ Weyl group is $192$, we find $1920/192=10$ cosets.
Using these cosets we can map the original parameter $(\overline{h}_1,\overline{h}_2,e_1,e_3,e_5,e_7)^{(2,2)}$ \eqref{heparameter} into other equivalent parameters.
In table \ref{D4vac} we display the $10$ parameters mapped by the cosets, the generators of the invariant $D_4$ Weyl groups and the one-dimensional subspaces invariant under these generators.

\begin{table}[!t]
\centering
\begin{tabular}{|c|c|c|c|}
\hline 
cosets & 
$(\overline{h}_{1},\overline{h}_{2},e_{1},e_{3},e_{5},e_{7})$ & $D_{4}$ symmetry & invariant subspace\\
\hline 
\hline 
$1,s_{3}s_{1}s_{2}s_{3}$ & $(q^{\pm1},q^{\mp1},1,1,1,1)$ & $\langle s_{3}s_{4}s_{3};s_{1},s_{2},s_{5}\rangle$ & $(h^{-1},h,1,1,1,1)$\\
\hline 
$s_{3},s_{1}s_{2}s_{3}$ & $(q^{\pm1},1,1,q^{\pm1},1,q^{\pm1})$ & $\langle s_{4};s_{1}s_{3}s_{1},s_{2}s_{3}s_{2},s_{5}\rangle$ & $(h^{-1},1,1,h^{-1},1,h^{-1})$\\
\hline 
$s_{4},s_{4}s_{3}s_{1}s_{2}s_{3}$ & $(1,q^{\mp1},q^{\mp1},1,q^{\mp1},1)$ & $\langle s_{3};s_{1},s_{2},s_{4}s_{5}s_{4}\rangle$ & $(1,h,h,1,h,1)$\\
\hline 
$s_{1}s_{3},s_{2}s_{3}$ & $(1,1,1,q^{\pm1},1,q^{\mp1})$ & $\langle s_{4};s_{3},s_{1}s_{2}s_{3}s_{2}s_{1},s_{5}\rangle$ & $(1,1,1,h^{-1},1,h)$\\
\hline 
$s_{5}s_{4},s_{5}s_{4}s_{3}s_{1}s_{2}s_{3}$ & $(1,1,q^{\pm1},1,q^{\mp1},1)$ & $\langle s_{3};s_{1},s_{2},s_{4}\rangle$ & $(1,1,h^{-1},1,h,1)$\\
\hline 
\end{tabular}
\caption{Different representative choices of the parameters for the $(2,2)$ model without rank deformations.
The parameters $(\overline{h}_1,\overline{h}_2,e_1,e_3,e_5,e_7)$ are obtained by acting the $10$ cosets of the $D_5$ Weyl group by the invariant $D_4$ Weyl group
 (where the upper/lower double-sign corresponds to the first/second coset respectively).
For the $D_{4}$ symmetry we first denote the root corresponding to the adjoint representation and then the other three corresponding to the three ${\bf 8}$ representations.}
\label{D4vac}
\end{table}

Similarly, we work for the $(1,1,1,1)$ model.
This time we find that, out of the $1920$ elements, there are $24$ elements that leave the parameter invariant.
We find that the whole $24$ elements are generated from the following four transformations
\begin{align}
s_{3},\quad s_{4},\quad s_{1}s_{3}s_{4}s_{5}s_{4}s_{3}s_{1},\quad s_{2}s_{3}s_{4}s_{5}s_{4}s_{3}s_{2}.
\end{align}
Since the latter two commute with the others it is clear that the invariant subgroup leaving the $(1,1,1,1)$ model is $A_{2}\times(A_{1})^{2}$ (see figure \ref{invalg} for the Dynkin diagram).
Compared with the analysis from the superconformal Chern-Simons matrix models where the invariant
subgroup was originally found to be $(A_{1})^{3}$ as reviewed in table \ref{a1^3},
the results do not coincide.
The reason is that in the study of the matrix models the invariant subgroup was found as the further deformation from the $(2,2)$ model and the subgroup was considered within the invariant subgroup of the $(2,2)$ model.
In fact, if we investigate the intersection of the $192$ elements of the invariant subgroup $D_{4}$ for the $(2,2)$ model and the $24$ elements of the invariant subgroup $A_{2}\times(A_{1})^{2}$, we find only $8$ elements containing the commuting elements 
\begin{align}
s_{3}s_{4}s_{3},\quad s_{1}s_{3}s_{4}s_{5}s_{4}s_{3}s_{1},\quad s_{2}s_{3}s_{4}s_{5}s_{4}s_{3}s_{2}.
\end{align}
Hence, the invariant subgroup is reduced to $(A_{1})^{3}$.
Alternatively, in \eqref{A2enhance} we have seen that the invariant subgroup of the $(1,1,1,1)$ model without rank deformations is enhanced to $A_2\times(A_1)^2$ accidentally.

\begin{figure}[!t]
\centering\includegraphics[scale=0.6,angle=90]{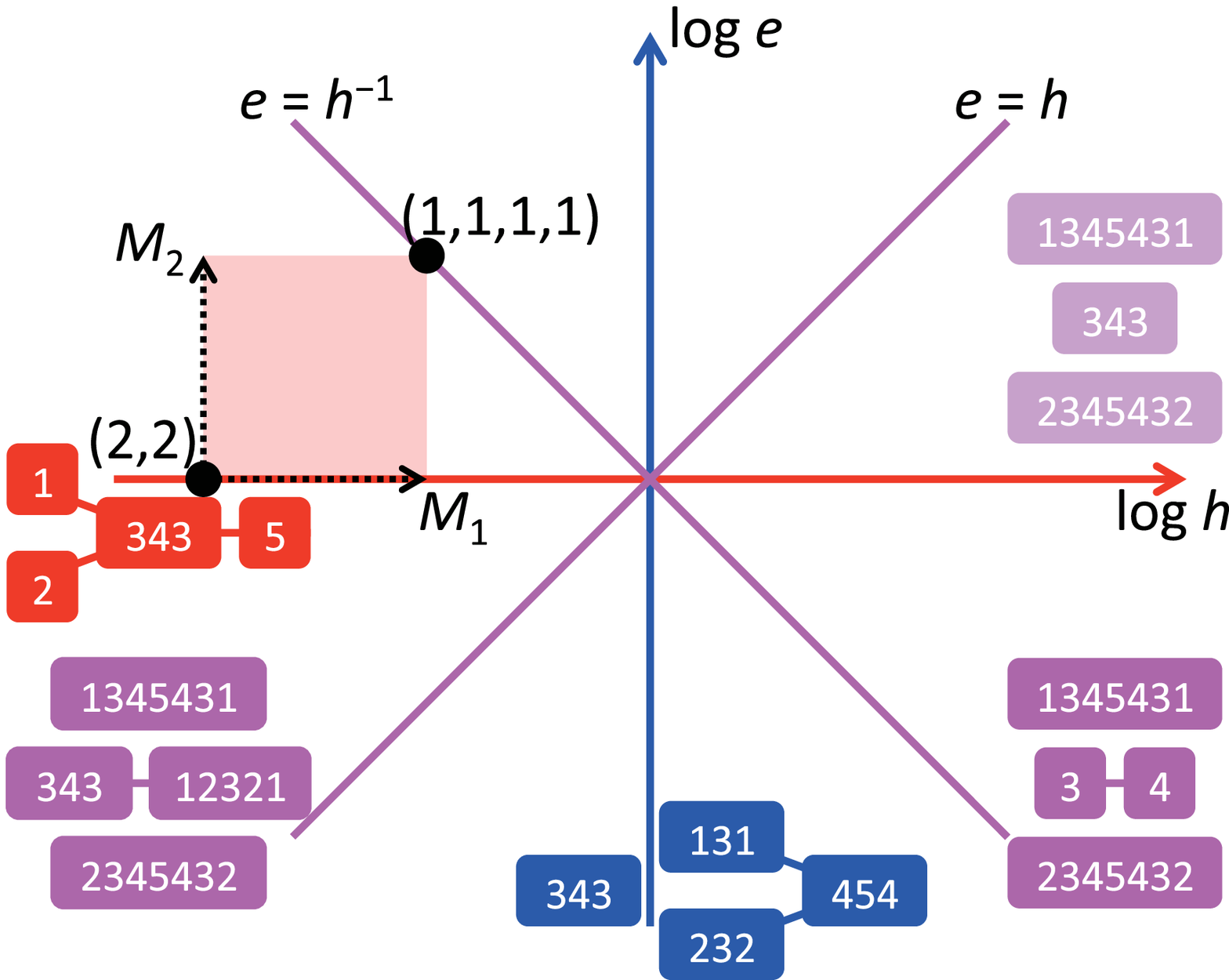}
\caption{The patterns of the symmetry breaking in the $(\log h,\log e)$ plane \eqref{space2}.}
\label{f=00003D1}
\end{figure}

Besides the $A_{1}$ symmetry $s_{3}s_{4}s_{3}$ requiring $h_{1}h_{2}=1$, since the actions of the extra two generators $s_{1}s_{3}s_{4}s_{5}s_{4}s_{3}s_{1}$ and $s_{2}s_{3}s_{4}s_{5}s_{4}s_{3}s_{2}$ are given by
\begin{align}
s_{1}s_{3}s_{4}s_{5}s_{4}s_{3}s_{1}:(h_{1},h_{2},e_{1},e_{3},e_{5};\alpha) & \mapsto\biggl(h_{1},h_{2},\frac{e_{1}e_{3}e_{5}}{h_{1}h_{2}},\frac{h_{1}h_{2}}{e_{5}},\frac{h_{1}h_{2}}{e_{3}};\frac{e_{3}e_{5}}{h_{1}h_{2}}\alpha\biggr),\nonumber \\
s_{2}s_{3}s_{4}s_{5}s_{4}s_{3}s_{2}:(h_{1},h_{2},e_{1},e_{3},e_{5};\alpha) & \mapsto\biggl(h_{1},h_{2},\frac{h_{1}h_{2}}{e_{3}},\frac{h_{1}h_{2}}{e_{1}},\frac{e_{1}e_{3}e_{5}}{h_{1}h_{2}};\alpha\biggr)\label{s34543}
\end{align}
the conditions $e_{1}=e_{3}^{-1}=e_{5}$ are further required.
Hence, the subspace enjoying the $(A_{1})^{3}$ symmetry is 
\begin{align}
\{(\overline{h}_{1},\overline{h}_{2},e_{1},e_{3},e_{5},e_7)
=(h^{-1}e,he^{-1},e^{-1},e,e^{-1},e)\}.
\label{space2}
\end{align}

As previously, we can use representatives of the cosets to map the parameter for the $(1,1,1,1)$ model without rank deformations into other parameters.
Since the order of the $A_2\times(A_1)^2$ invariant subgroup is $24$, we have $1920/24=80$ cosets and hence $80$ equivalent parameters for the same model.

Since it was known that the $(2,2)$ model and the $(1,1,1,1)$ model are connected through rank deformations as studied carefully in \cite{MNN}, it is natural to expect that they are also connected in the parameter space.
Intending to understand better ``the moduli space of the M2-branes'', we concentrate on the subspace \eqref{space2} and study the invariant subgroup at each point.
We find that the symmetry enhances at certain linear subspaces as depicted in figure \ref{f=00003D1}.

We can identify the one-dimensional subspace \eqref{space1} spanned by $h$ as the $M_{\text{I}}$  deformation space of the $(2,2)$ model and the two-dimensional subspace \eqref{space2} as the $(M_{\text{I}},M_{\text{II}})$ deformation space because of the correspondence of the symmetries. Furthermore, if we take the fact into account that the $(2,2)$ model with $(M_{\text{I}},M_{\text{II}})=(k/2,k/2)$ is equal to the $(1,1,1,1)$ model without rank deformations as explained above \eqref{kalher1111}, we can tentatively identify the correspondence between parameters of the curve and parameters of the $(2,2)$ model as
\begin{align}
h=e^{2\pi i(M_{\text{I}}-k)},\quad
e=e^{2\pi iM_{\text{II}}}.
\end{align}
The result of the rank deformations in \cite{MNN} seems consistent with our current analysis.
To really understand the rank deformations, however, we need some further clarifications which is beyond the scope of the present work \cite{KMN2}.

Finally let us turn to the $(2,1)$ model.
There are $24$ elements containing 
\begin{align}
s_{1},\quad s_{2},\quad s_{3}s_{4}s_{3},
\end{align}
which leave \eqref{heparameter21} invariant.
These three elements are part of (\ref{D4generators}), so the invariant subgroup for the $(2,1)$ model is $A_{3}$, which again accords with the analysis on the matrix model side.

\begin{figure}[!t]
\centering\includegraphics[scale=0.6,angle=-90]{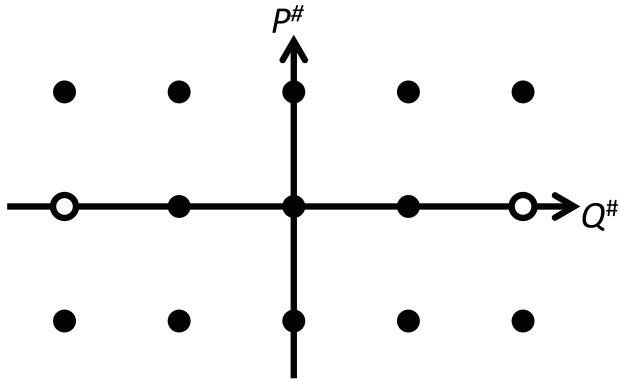}
\caption{The Newton polygon of the $E_{7}$ algebra.}
\label{e7np}
\end{figure}

\section{Degenerate curve}\label{sec_SEC}

\begin{figure}[!t]
\centering\includegraphics[scale=0.6,angle=90]{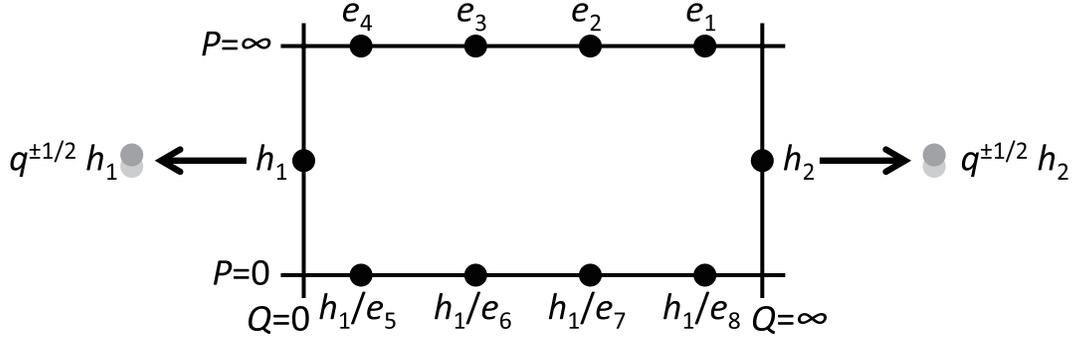}
\caption{The asymptotic values of the $E_{7}$ del Pezzo curve.}
\label{e7asymp}
\end{figure}

Let us turn to the $E_{7}$ del Pezzo curve.
Classically, the $E_{7}$ del Pezzo curve is realized as a linear combination of
\begin{align}
{Q}^{m}{P}^{n},\quad m=-2,-1,0,1,2,\quad n=-1,0,1.
\end{align}
See figure \ref{e7np} for the Newton polygon.
This realization of the $E_{7}$ del Pezzo curve appeared in \cite{KimYagi} following the proposal of utilizing the degenerate genus in \cite{BBT}.
Note that the coefficients of these operators are not all independent, otherwise the number of inner points in the Newton polygon indicating that the genus is three.
To reduce the genus to one, classically we require the curve to be singular at $(Q,P)=(0,h_{1})$ and $(Q,P)=(\infty,h_{2})$.
Since the singular point of an algebraic curve $H(Q,P)=0$ at $(Q_{0},P_{0})$ is defined by 
\begin{align}
H(Q_{0},P_{0})=0,\quad
\frac{\partial H}{\partial Q}(Q_{0},P_{0})=0,\quad
\frac{\partial H}{\partial P}(Q_{0},P_{0})=0,
\label{SingCon}
\end{align}
the requirements we have imposed become the conditions that the quadratic polynomials of $P$ at $Q^{2}$ and $Q^{1}$ respectively have a double root and a single root at $h_{2}$ and those at $Q^{-2}$ and $Q^{-1}$ respectively have a double root and a single root at $h_{1}$.
For other asymptotical values, we set 
\begin{align}
(e_{1},\infty),(e_{2},\infty),(e_{3},\infty),(e_{4},\infty),(h_{1}e_{5}^{-1},0),(h_{1}e_{6}^{-1},0),(h_{1}e_{7}^{-1},0),(h_{1}e_{8}^{-1},0).
\end{align}
See figure \ref{e7asymp} for the asymptotical values.
Again from the Vieta's formulas on products of roots, the parameters satisfy
\begin{align}
\prod_{i=1}^{8}e_{i}=h_{1}^{2}h_{2}^{2}.
\end{align}

\begin{figure}[!t]
\centering\includegraphics[scale=0.6,angle=-90]{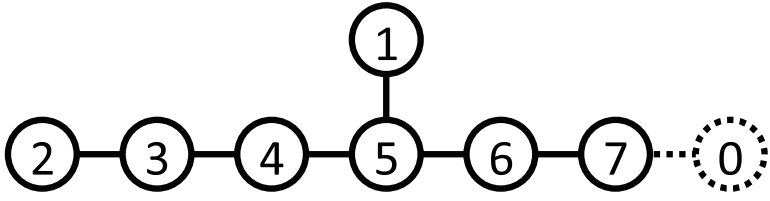}
\caption{The Dynkin diagram of the $E_{7}$ algebra.}
\label{e7dd}
\end{figure}

Then we find that if we define $s_{1}$ to be the exchange of the two singular asymptotical points
\begin{align}
s_{1}:h_{1}\leftrightarrow h_{2},
\end{align}
generated by the canonical transformation
\begin{align}
Q'=(P-h_{1})^{-1}Q(P-h_{2}),\quad P'=P,\label{canonical}
\end{align}
along with
\begin{align}
&s_{2}:e_{4}\leftrightarrow e_{3},\quad
s_{3}:e_{3}\leftrightarrow e_{2},\quad
s_{4}:e_{2}\leftrightarrow e_{1},\quad
s_{5}:e_{1}\leftrightarrow h_{1}e_{5}^{-1},\nonumber\\
&s_{6}:h_{1}e_{5}^{-1}\leftrightarrow h_{1}e_{6}^{-1},\quad
s_{7}:h_{1}e_{6}^{-1}\leftrightarrow h_{1}e_{7}^{-1},
\end{align}
these actions generate the whole $E_{7}$ Weyl group whose Dynkin diagram is given in figure \ref{e7dd}.

Now let us turn to quantum curves.
Our working hypothesis for the condition of the degeneracy \eqref{SingCon} for the quantum curves is that the relative coefficients are determined so that the quantum deformation of the transformation \eqref{canonical} is again the symmetry of the curve.
Then, for the transformation in the positive or negative quadratic terms to work we need to split
the asymptotic double roots $h_{1}$ and $h_{2}$ respectively into $q^{\pm\frac{1}{2}}h_{1}$ and $q^{\pm\frac{1}{2}}h_{2}$ and consider the curve specified by
\begin{align}
&\widehat{H}/\alpha=\widehat{Q}^{2}(\widehat{P}-q^{-\frac{1}{2}}h_{2})(\widehat{P}-q^{\frac{1}{2}}h_{2})\widehat{P}^{-1}\nonumber\\
&\quad-\widehat{Q}(\widehat{P}-q^{-\frac{1}{2}}h_{2})\bigl((e_{1}+e_{2}+e_{3}+e_{4})\widehat{P}-q^{\frac{1}{2}}h_{1}h_{2}(e_{5}^{-1}+e_{6}^{-1}+e_{7}^{-1}+e_{8}^{-1})\bigr)\widehat{P}^{-1}\nonumber\\
&\quad+\Bigl\{ (e_{1}e_{2}+e_{1}e_{3}+e_{1}e_{4}+e_{2}e_{3}+e_{2}e_{4}+e_{3}e_{4})\widehat{P}
+E/\alpha\nonumber\\
&\qquad+h_{1}^{2}h_{2}^{2}(e_{5}^{-1}e_{6}^{-1}+e_{5}^{-1}e_{7}^{-1}+e_{5}^{-1}e_{8}^{-1}+e_{6}^{-1}e_{7}^{-1}+e_{6}^{-1}e_{8}^{-1}+e_{7}^{-1}e_{8}^{-1})\widehat{P}^{-1}\Bigr\}\nonumber\\
&\quad-e_{1}e_{2}e_{3}e_{4}Q^{-1}(P-q^{\frac{1}{2}}h_{1})\bigl((e_{1}^{-1}+e_{2}^{-1}+e_{3}^{-1}+e_{4}^{-1})\widehat{P}-q^{-\frac{1}{2}}(e_{5}+e_{6}+e_{7}+e_{8})\bigr)\widehat{P}^{-1}\nonumber\\
&\quad+e_{1}e_{2}e_{3}e_{4}\widehat{Q}^{-2}(\widehat{P}-q^{\frac{1}{2}}h_{1})(\widehat{P}-q^{-\frac{1}{2}}h_{1})\widehat{P}^{-1}.
\label{generalE7curve}
\end{align}
Instead we can display the same curve by listing each order of ${\widehat P}$, as
\begin{align}
&\widehat{H}/\alpha
=(\widehat{Q}-e_{1})(\widehat{Q}-e_{2})(\widehat{Q}-e_{3})(\widehat{Q}-e_{4})
\widehat{Q}^{-2}\widehat{P}\nonumber\\
&\quad+\Bigl\{-(q^{\frac{1}{2}}+q^{-\frac{1}{2}})h_{2}\widehat{Q}^{2}
+h_{2}\bigl[q^{-\frac{1}{2}}(e_{1}+e_{2}+e_{3}+e_{4})+q^{\frac{1}{2}}h_{1}(e_{5}^{-1}+e_{6}^{-1}+e_{7}^{-1}+e_{8}^{-1})\bigr]\widehat{Q}\nonumber\\
&\qquad+E/\alpha+e_{1}e_{2}e_{3}e_{4}\bigl[q^{-\frac{1}{2}}(e_{5}+e_{6}+e_{7}+e_{8})+q^{\frac{1}{2}}h_{1}(e_{1}^{-1}+e_{2}^{-1}+e_{3}^{-1}+e_{4}^{-1})\bigr]\widehat{Q}^{-1}
\nonumber\\
&\qquad
-(q^{\frac{1}{2}}+q^{-\frac{1}{2}})h_{1}e_{1}e_{2}e_{3}e_{4}\widehat{Q}^{-2}\Bigr\}\nonumber\\
&\quad+h_{2}^{2}(\widehat{Q}-h_{1}e_{5}^{-1})(\widehat{Q}-h_{1}e_{6}^{-1})(\widehat{Q}-h_{1}e_{7}^{-1})(\widehat{Q}-h_{1}e_{8}^{-1})\widehat{Q}^{-2}\widehat{P}^{-1}.
\end{align}

Let us turn to the Weyl symmetry of the curve.
For $s_{1}$, we consider the similarity transformation
\begin{align}
\widehat{Q}'=(\widehat{P}-q^{\frac{1}{2}}h_{1})^{-1}\widehat{Q}(\widehat{P}-q^{-\frac{1}{2}}h_{2}),\quad\widehat{P}'=\widehat{P},
\label{quantumcanonical}
\end{align}
which also implies
\begin{align}
\widehat{Q}'^{2}=(q\widehat{P}-q^{\frac{1}{2}}h_{1})^{-1}(\widehat{P}-q^{\frac{1}{2}}h_{1})^{-1}\widehat{Q}^{2}(\widehat{P}-q^{-\frac{1}{2}}h_{2})(q^{-1}\widehat{P}-q^{-\frac{1}{2}}h_{2}).
\end{align}
After combining with the rescaling
\begin{align}
\widehat{Q}''=q^{-1}\widehat{Q}',\quad\widehat{P}''=\widehat{P}'
\end{align}
we find that $s_{1}$ transforms the parameters as
\begin{align}
s_{1}:(h_{1},h_{2},e_{1},e_{2},e_{3},e_{4},e_{5},e_{6},e_{7},e_{8};\alpha)\mapsto(q^{-2}h_{2},q^{2}h_{1},e_{1},e_{2},e_{3},e_{4},e_{5},e_{6},e_{7},e_{8};\alpha)
\end{align}
The other transformations are parallel to the previous studies in the $D_{5}$ case.

As in the $D_{5}$ case, if we use the degrees of freedom of the rescaling of $\widehat{Q}$ and $\widehat{P}$ \eqref{rescaling} to fix the gauge
\begin{align}
e_{4}=e_{8}=1,
\end{align}
and drop $e_7$ by using the constraint
\begin{align}
e_{7}=\frac{h_{1}^{2}h_{2}^{2}}{e_{1}e_{2}e_{3}e_{5}e_{6}},
\end{align}
we find that the transformations are given by
\begin{align}
 & s_{1}:(h_{1},h_{2},e_{1},e_{2},e_{3},e_{5},e_{6};\alpha)\mapsto\biggl(\frac{h_{2}}{q^{2}},q^{2}h_{1},e_{1},e_{2},e_{3},e_{5},e_{6};\alpha\biggr)\nonumber\\
 & s_{2}:(h_{1},h_{2},e_{1},e_{2},e_{3},e_{5},e_{6};\alpha)\mapsto\biggl(\frac{h_{1}}{e_{3}},\frac{h_{2}}{e_{3}},\frac{e_{1}}{e_{3}},\frac{e_{2}}{e_{3}},\frac{1}{e_{3}},e_{5},e_{6};e_{3}^{3}\alpha\biggr),\nonumber\\
 & s_{3}:(h_{1},h_{2},e_{1},e_{2},e_{3},e_{5},e_{6};\alpha)\mapsto(h_{1},h_{2},e_{1},e_{3},e_{2},e_{5},e_{6};\alpha),\nonumber\\
 & s_{4}:(h_{1},h_{2},e_{1},e_{2},e_{3},e_{5},e_{6};\alpha)\mapsto(h_{1},h_{2},e_{2},e_{1},e_{3},e_{5},e_{6};\alpha),\nonumber\\
 & s_{5}:(h_{1},h_{2},e_{1},e_{2},e_{3},e_{5},e_{6};\alpha)\mapsto\biggl(h_{1},\frac{qh_{1}h_{2}}{e_{1}e_{5}},\frac{qh_{1}}{e_{5}},e_{2},e_{3},\frac{qh_{1}}{e_{1}},e_{6};\frac{e_{1}e_{5}}{qh_{1}}\alpha\biggr),\nonumber\\
 & s_{6}:(h_{1},h_{2},e_{1},e_{2},e_{3},e_{5},e_{6};\alpha)\mapsto(h_{1},h_{2},e_{1},e_{2},e_{3},e_{6},e_{5};\alpha)\nonumber\\
 & s_{7}:(h_{1},h_{2},e_{1},e_{2},e_{3},e_{5},e_{6};\alpha)\mapsto\biggl(h_{1},h_{2},e_{1},e_{2},e_{3},e_{5},\frac{h_{1}^{2}h_{2}^{2}}{e_{1}e_{2}e_{3}e_{5}e_{6}};\alpha\biggr).\label{WeylE7}
\end{align}
Note again that, if we introduce $\overline{h}_{1}$ and $\overline{h}_{2}$ as in \eqref{hbar},
\begin{align}
\overline{h}_1=qh_1,\quad\overline{h}_2=q^{-1}h_2,
\end{align}
we can absorb the quantum deformation parameter $q$ completely in \eqref{WeylE7}.

As in the $D_5$ case, we find that the Weyl symmetries are realized as standard Weyl actions.
In this case, root vectors and fundamental weight vectors are
\begin{align}
&\alpha_{1}=(1,-1,0,0,0,0,0),\quad\omega_1=(2,1,0,0,0,2,2),\nonumber\\
&\alpha_{2}=(-1,-1,-1,-1,-2,0,0),\quad\omega_2=(0,0,-1,-1,-1,1,1),\nonumber\\
&\alpha_{3}=(0,0,0,-1,1,0,0),\quad\omega_3=(1,1,-1,-1,0,2,2),\nonumber\\
&\alpha_{4}=(0,0,-1,1,0,0,0),\quad\omega_4=(2,2,-1,0,0,3,3),\nonumber\\
&\alpha_{5}=(0,1,1,0,0,1,0),\quad\omega_5=(3,3,0,0,0,4,4),\nonumber\\
&\alpha_{6}=(0,0,0,0,0,-1,1),\quad\omega_6=(2,2,0,0,0,2,3),\nonumber\\
&\alpha_{7}=(0,0,0,0,0,0,-1),\quad\omega_7=(1,1,0,0,0,1,1),
\end{align}
where we have represented the parameters by $(\log\overline{h}_1,\log\overline{h}_2,\log e_1,\log e_2,\log e_3,\log e_5,\log e_6)$ as in the $D_5$ case.

In the study of the $(2,1,2,1)$ model without rank deformations, it was found that the model falls into the class of the $E_{7}$ del Pezzo curve with the symmetry broken to $D_{5}\times A_{1}$.
As in the previous case we can identify the parameters of the $(2,1,2,1)$ model
\begin{align}
&\widehat{H}=(\widehat{Q}^{\frac{1}{2}}+\widehat{Q}^{-\frac{1}{2}})^{2}(\widehat{P}^{\frac{1}{2}}+\widehat{P}^{-\frac{1}{2}})(\widehat{Q}^{\frac{1}{2}}+\widehat{Q}^{-\frac{1}{2}})^{2}(\widehat{P}^{\frac{1}{2}}+\widehat{P}^{-\frac{1}{2}})\nonumber\\
&=q^{-\frac{1}{2}}\widehat{Q}^{2}\widehat{P}+2(1+q^{-\frac{1}{2}})\widehat{Q}\widehat{P}+(q^{\frac{1}{2}}+4+q^{-\frac{1}{2}})\widehat{P}+2(q^{\frac{1}{2}}+1)\widehat{Q}^{-1}\widehat{P}+q^{\frac{1}{2}}\widehat{Q}^{-2}\widehat{P}\nonumber\\
&\quad+(q^{\frac{1}{2}}+q^{-\frac{1}{2}})\widehat{Q}^{2}+2(q^{\frac{1}{2}}+2+q^{-\frac{1}{2}})\widehat{Q}+2(q^{\frac{1}{2}}+4+q^{-\frac{1}{2}})\nonumber\\
&\qquad\qquad\qquad\qquad+2(q^{\frac{1}{2}}+2+q^{-\frac{1}{2}})\widehat{Q}^{-1}+(q^{\frac{1}{2}}+q^{-\frac{1}{2}})\widehat{Q}^{-2}\nonumber\\
&\quad+q^{\frac{1}{2}}\widehat{Q}^{2}\widehat{P}^{-1}+2(q^{\frac{1}{2}}+1)\widehat{Q}\widehat{P}^{-1}+(q^{\frac{1}{2}}+4+q^{-\frac{1}{2}})\widehat{P}^{-1}\nonumber\\
&\qquad\qquad\qquad\qquad+2(1+q^{-\frac{1}{2}})\widehat{Q}^{-1}\widehat{P}^{-1}+q^{-\frac{1}{2}}\widehat{Q}^{-2}\widehat{P}^{-1},
\end{align}
as
\begin{align}
(h_{1},h_{2},e_{1},e_{2},e_{3},e_{5},e_{6},e_{7})^{(2,1,2,1)}=(1,q,q^{\frac{1}{2}},q^{\frac{1}{2}},1,q^{\frac{1}{2}},q^{\frac{1}{2}},1).
\end{align}

\begin{figure}[!t]
\centering\includegraphics[scale=0.6,angle=-90]{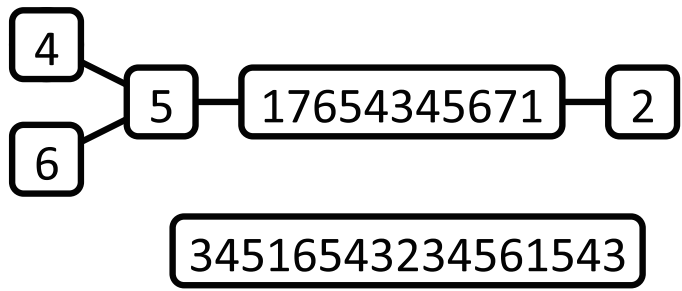}
\caption{The Dynkin diagram of the $D_{5}\times A_{1}$ subalgebra within the original $E_7$ algebra, which preserves the $(2,1,2,1)$ model without rank deformations.}
\label{e7inv}
\end{figure}

Then we can ask again which elements of the $E_{7}$ Weyl group generated by $s_{1},s_{2},s_{3},\cdots,s_{7}$ preserve this parameter and what group those elements form.
We again generate all elements of the $E_{7}$ Weyl group by using a computer, then we find that the answer is $3840$ elements generated by
\begin{align}
s_4,\quad s_5,\quad s_6,\quad s_1s_7s_6s_5s_4s_3s_4s_5s_6s_7s_1,\quad s_2,
\end{align}
along with the commuting element $s_3s_4s_5s_1s_6s_5s_4s_3s_2s_3s_4s_5s_6s_1s_5s_4s_3$.
See figure \ref{e7inv} for the Dynkin diagram for these elements.
The former five elements generate the Weyl group of $D_{5}$ while the latter is $A_{1}$.
The result matches again with the study from the superconformal Chern-Simons matrix model.

Since the $E_{7}$ Weyl group has $2903040$ elements in total, it is not easy to generate the elements without strategies.
We shall explain how we have generated them in appendix \ref{e7weyl}.

\section{Conclusion}

In this paper we have studied the symmetry breaking of the quantum curves.
We first find that the symmetry for the classical algebraic curve given in the Weyl group of the exceptional algebra is promoted to that for the quantum curve in our definition.
We then fix the values of the parameters to those of the superconformal Chern-Simons matrix models and study the symmetry breaking patterns for these values.

The main motivation of our work is to reproduce the symmetry breaking patterns we found previously for the superconformal Chern-Simons matrix models.
After the reproduction, we continue to study the breaking patterns of other values, partially expecting that this gives ``the moduli space of the M2-branes''.
We find that the moduli space does not change from the classical limit $q\to 1$ and conjecture that this is the case as well for other curves.
It is interesting to find that the whole moduli space of the M2-branes enjoys a generalization of the Weyl group of the exceptional algebra.

Previously the correspondence between the superconformal Chern-Simons matrix models and the algebraic curves was mainly studied from the analytical viewpoint.
We believe that, along with \cite{MNY}, our computation in terms of the symmetry breaking of the Weyl group has opened up a new avenue to understand better the correspondence.
We shall list several further directions.

First, our method is applicable to many generalizations and the study in these directions may lead to many clarifications of the correspondence.
As well as the grand canonical ensembles of the matrix models constructed from other ${\cal N}=4$ U$(N)^{r+1}$ superconformal Chern-Simons theories for the $\widehat A_r$ quiver diagram with \eqref{sN4}, those constructed from the ${\cal N}=3$ theories with the same field contents (\eqref{sN4} with $\{s_a\}_{a=1}^{r+1}$ being arbitrary integers) also take the form of the Fredholm determinant \eqref{fredholm}.
In these cases ${\widehat H}$ is given as a product of $2\cosh({{\widehat q}-s_a{\widehat p}})/{2}$, and hence is expanded by finite terms of ${\widehat Q}^m{\widehat P}^n$ with $(m,n)\in\mathbb{Z}^2$.
In general, the curve is a higher-genus generalization of the del Pezzo curves we have considered in this paper.
It is interesting to study the higher-genus generalizations and compare with the results in \cite{CGM}.

Secondly, so far in this paper we have mainly restricted our studies to the case without rank deformations.
To fully understand the moduli space of the M2-branes we need to proceed to the rank deformations.
For the rank deformations, however, we encounter several interesting new points to be clarified which we would like to study more carefully and report in our future work \cite{KMN2}.

Thirdly, the rank deformations of the $(2,1,2,1)$ model have another interesting aspect.
If we consider the $(2,1,2,1)$ model with general choice of six ranks, the whole moduli space would also include the $(4,2)$ model, whose curve is not degenerate according to the definition of degenerate quantum curve in section \ref{sec_SEC}.
It would be interesting to study the rank deformations of the $(2,1,2,1)$ model and identify the special class which keeps the degeneracy of the $E_7$ curve.

Fourthly, it would be nice to establish the quantum notion of degenerate curves for general type of singularities.
In the case of the classical $E_7$ curve, there is a pair of singularities at $Q=0$ and $Q=\infty$, which are exchanged by the symmetry $s_1$ \eqref{canonical}.
Hence we can define the degeneracy condition for the quantum curve by requiring that $s_1$, now the similarity transformation of the quantum operators \eqref{quantumcanonical}, remains to be the symmetry of the quantum curves.
This strategy does not work for general un-paired singularities.
Nevertheless, we notice that, if we introduce the $q$-derivative as $d^q_xf(x)=(f(qx)-f(x))/(qx-x)$, the $E_7$ curve \eqref{generalE7curve} satisfies $H(0,q^{-1/2}h_2)=d^q_QH(0,q^{-1/2}h_2)=d^q_PH(0,q^{-1/2}h_2)=0$ at the singularity on $Q=0$ (and the same condition for $Q=\infty$).
This is analogous to the degeneracy condition for the classical curves \eqref{SingCon}, hence might be a good starting point.
Once we know the definition of quantum degeneracy, it would be possible to study the models corresponding to the $E_6$ curve or the $E_8$ curve as well, which can be obtained by starting from the higher genus rectangular curve and tuning the parameters so that the curve is singular and all but a single genus are degenerate, similar to the case of the $E_7$ curve \cite{KimYagi}.

Fifthly, the relation to the $q$-deformed Painlev\'e equation is another interesting direction.
It was shown in \cite{BGT} that the grand canonical ensemble of the ABJM matrix model satisfies the $q$-deformed Painlev\'e equation.
On the other hand, the Weyl symmetries we have studied in this paper are also known to be the symmetries of the $q$-deformed Painlev\'e equations \cite{KNY}.
Hence, our studies of the relation between the superconformal Chern-Simons matrix models and the Weyl symmetries should be connected via the $q$-deformed Painlev\'e equations.
The integrable structure of the superconformal Chern-Simons matrix models would be made clearer from the studies.
Recent developments between matrix models and Painlev\'e equations \cite{GG,IOY} may be helpful in studying the generalizations.

Finally, it is also interesting to study the connection to five-dimensional gauge theories.
Historically the Newton polygons and the dual asymptotic values played an important role in studying five-dimensional gauge theories constructed from the $(p,q)$5-brane webs \cite{AHK}.
The superconformal Chern-Simons matrix models studied in this paper can also be realized by the type IIB brane setups consisting of D3-branes spanning between NS5-branes and $(1,k)$5-branes, where the rank of the gauge group is given by the number of D3-branes on each segment.
We hope to clarify the relation between the five-dimensional gauge theories and the three-dimensional gauge theories.

\appendix

\section{Weyl group}\label{e7weyl}

\begin{table}[!t]
\begin{center}
\begin{tabular}{c|c|c|c|c|c}
$E_3=A_{2}\times A_{1}$ & $E_4=A_{4}$ & $E_5=D_{5}$ & $E_{6}$ & $E_{7}$ & $E_{8}$\\
\hline 
$12$ & $192$ & $1920$ & $51840$ & $2903040$ & $696729600$ \\
\end{tabular}
\end{center}
\caption{The order of the Weyl group for each exceptional algebra $\#W(E_n)$.}
\label{card}
\end{table}

\begin{table}[!t]
\begin{center}
\begin{tabular}{c|c|c|c|c}
$A_{4}/(A_{2}\times A_{1})$ & $D_{5}/A_{4}$ & $E_{6}/D_{5}$ & $E_{7}/E_{6}$ & $E_{8}/E_{7}$\\
\hline
$16$ & $10$ & $27$ & $56$ & $240$ \\
\end{tabular}
\end{center}
\caption{The number of cosets of two Weyl groups $\#\bigl(W(E_n)/W(E_{n-1})\bigr)$.}
\label{cardcoset}
\end{table}

In this appendix we comment on the computation of the Weyl group of the exceptional algebra, when the order, the number of the elements, is large such as $E_6,E_7,E_8$.
See table \ref{card} for the order of the Weyl group for the exceptional algebra $W(E_n)$ where we denote the Weyl group of the Lie algebra $G$ by $W(G)$.

A first trial would be to collect all of different elements of the Weyl group from the simple roots by multiplying them one by one to see whether the transformation is new or not.
Namely, we prepare a set of elements obtained so far and try to generate a new element of the Weyl group by multiplying the simple roots to those in the original set.
If the transformation is new, we add the new element to the original set.
Otherwise we forget it and proceed to the next multiplication.
This is valid for the $D_5$ Weyl group with only $1920$ elements in total, though for the $E_{7}$ Weyl group with $2903040$ elements this method is very time-consuming.
The article \cite{Mashimo} is helpful for us to improve this situation.

\begin{table}[!t]
\begin{center}
\begin{tabular}{c|l}
step & representatives\\
\hline 
$0$ & $1$\\
$1$ & $s_{3}$\\
$2$ & $s_{3}s_{4}$\\
$3$ & $s_{3}s_{4}s_{5}$\\
$4$ & $s_{3}s_{4}s_{5}s_{1},s_{3}s_{4}s_{5}s_{6}$\\
$5$ & $s_{3}s_{4}s_{5}s_{1}s_{6},s_{3}s_{4}s_{5}s_{6}s_{7}$\\
$6$ & $s_{3}s_{4}s_{5}s_{1}s_{6}s_{5},s_{3}s_{4}s_{5}s_{1}s_{6}s_{7}$\\
$7$ & $s_{3}s_{4}s_{5}s_{1}s_{6}s_{5}s_{4},s_{3}s_{4}s_{5}s_{1}s_{6}s_{5}s_{7}$\\
$8$ & $s_{3}s_{4}s_{5}s_{1}s_{6}s_{5}s_{4}s_{3},s_{3}s_{4}s_{5}s_{1}s_{6}s_{5}s_{4}s_{7},s_{3}s_{4}s_{5}s_{1}s_{6}s_{5}s_{7}s_{6}$\\
$9$ & $s_{3}s_{4}s_{5}s_{1}s_{6}s_{5}s_{4}s_{3}s_{7},s_{3}s_{4}s_{5}s_{1}s_{6}s_{5}s_{4}s_{7}s_{6}$\\
$10$ & $s_{3}s_{4}s_{5}s_{1}s_{6}s_{5}s_{4}s_{3}s_{7}s_{6},s_{3}s_{4}s_{5}s_{1}s_{6}s_{5}s_{4}s_{7}s_{6}s_{5}$\\
$11$ & $s_{3}s_{4}s_{5}s_{1}s_{6}s_{5}s_{4}s_{3}s_{7}s_{6}s_{5},s_{3}s_{4}s_{5}s_{1}s_{6}s_{5}s_{4}s_{7}s_{6}s_{5}s_{1}$\\
$12$ & $s_{3}s_{4}s_{5}s_{1}s_{6}s_{5}s_{4}s_{3}s_{7}s_{6}s_{5}s_{1},s_{3}s_{4}s_{5}s_{1}s_{6}s_{5}s_{4}s_{3}s_{7}s_{6}s_{5}s_{4}$\\
$13$ & $s_{3}s_{4}s_{5}s_{1}s_{6}s_{5}s_{4}s_{3}s_{7}s_{6}s_{5}s_{1}s_{4}$\\
$14$ & $s_{3}s_{4}s_{5}s_{1}s_{6}s_{5}s_{4}s_{3}s_{7}s_{6}s_{5}s_{1}s_{4}s_{5}$\\
$15$ & $s_{3}s_{4}s_{5}s_{1}s_{6}s_{5}s_{4}s_{3}s_{7}s_{6}s_{5}s_{1}s_{4}s_{5}s_{6}$\\
$16$ & $s_{3}s_{4}s_{5}s_{1}s_{6}s_{5}s_{4}s_{3}s_{7}s_{6}s_{5}s_{1}s_{4}s_{5}s_{6}s_{7}$\\
\end{tabular}
\end{center}
\caption{The $27$ cosets of the $D_{5}$ Weyl group (generated by $s_{1},s_{4},s_{5},s_{6},s_{7}$) in the $E_{6}$ Weyl group (generated by $s_{1},s_{3},s_{4},s_{5},s_{6},s_{7}$).}
\label{d5e6}
\end{table}

\begin{table}[!t]
\begin{center}
\begin{tabular}{c|l}
step & representatives\\
\hline 
$1$ & $2$\\
$2$ & $23$\\
$3$ & $234$\\
$4$ & $2345$\\
$5$ & $23451,23456$\\
$6$ & $234516,234567$\\
$7$ & $2345165,2345167$\\
$8$ & $23451654,23451657$\\
$9$ & $234516543,234516547,234516576$\\
$10$ & $2345165432,2345165437,2345165476$\\
$11$ & $23451654327,23451654376,23451654765$\\
$12$ & $234516543276,234516543765,234516547651$\\
$13$ & $2345165432765,2345165437651,2345165437654$\\
$14$ & $23451654327651,23451654327654,23451654376514$\\
$15$ & $234516543276514,234516543276543,234516543765145$\\
$16$ & $2345165432765143,2345165432765145,2345165437651456$\\
$17$ & $23451654327651435,23451654327651456,23451654376514567$\\
$18$ & $234516543276514354,234516543276514356,234516543276514567$\\
$19$ & $2345165432765143546,2345165432765143567$\\
$20$ & $23451654327651435465,23451654327651435467$\\
$21$ & $234516543276514354651,234516543276514354657$\\
$22$ & $2345165432765143546517,2345165432765143546576$\\
$23$ & $23451654327651435465176$\\
$24$ & $234516543276514354651765$\\
$25$ & $2345165432765143546517654$\\
$26$ & $23451654327651435465176543$\\
$27$ & $234516543276514354651765432$\\
\end{tabular}
\end{center}
\caption{The $56$ cosets of the $E_{6}$ Weyl group (generated by $s_{1},s_{3},s_{4},s_{5},s_{6},s_{7}$) in the $E_{7}$ Weyl group (generated by $s_{1},s_{2},s_{3},s_{4},s_{5},s_{6},s_{7}$).
We abbreviate $s_{1},s_{2},s_{3},s_{4},s_{5},s_{6},s_{7}$ as $1,2,3,4,5,6,7$ for simplicity.}
\label{e6e7}
\end{table}

The main time-consuming process would be to judge whether the generated element is new or not.
In fact, for the final step of the $E_7$ Weyl group, we need to multiply $7$ elements to the existing $2903040-1$ elements in the set to find out only one, the longest element, by comparing the generated $(2903040-1)\times 7$ elements with each of the $2903040-1$ elements.
To improve the situation, it is nice to consider cosets.
Namely, if we need to study the Weyl group of $G$, we can start from the Weyl group of a subalgebra $H(\subset G)$ and consider only the cosets in $W(G)/W(H)$.
When we generate the $E_6$ Weyl group from the $D_5$ Weyl group the number of the cosets $W(E_6)/W(D_5)$ is $27$, while when we generate the $E_7$ Weyl group from the $E_6$ Weyl group the number of the cosets $W(E_7)/W(E_6)$ is $56$.
See table \ref{cardcoset} for the number of the cosets of two Weyl groups for the subsequent exceptional algebras.
Then, each time we find out a new coset in $W(G)/W(H)$, as a bonus, we obtain $\#W(H)$ new elements in the Weyl group $W(G)$.
We have followed this process to find out the $E_7$ Weyl group.
Namely, starting from the $D_{5}$ Weyl group generated by $s_{1},s_{4},s_{5},s_{6},s_{7}$, to find out the $E_{6}$ Weyl group generated by $s_{1},s_{3},s_{4},s_{5},s_{6},s_{7}$ all we have to do is to find out the $27$ cosets.
(The numbering of the simple roots is the same as in figure \ref{e7dd}.)
Each time we find out one element in the coset, we can generate $1920$ elements in the original $E_{6}$ Weyl group.
See table \ref{d5e6} for the representatives of the cosets $W(E_6)/W(D_5)$.
After generating the $E_{6}$ Weyl group, to find out the $E_{7}$ Weyl group, we only need to find out the $56$ cosets, which can be done similarly.
See table \ref{e6e7} for the representatives of the cosets $W(E_7)/W(E_6)$.
This process saves a lot of time, though the computation still takes several hours on a decent laptop computer.

We can further improve the computation.
Namely, although we have reduced the computation by considering the coset, in the final step for $E_7$ we still need to generate $(56-1)\times 7$ elements and compare them with the existing $2903040-51840$ elements.
It is nice if we can compare the generated elements only with those in the cosets $W(G)/W(H)$.

The main idea in \cite{Mashimo} is to restrict the consideration to a Weyl chamber of $W(H)$.
Namely, by changing our transformations into the standard Weyl actions, we can choose the representatives of the cosets in the Weyl chamber.
Since we only consider those representatives, if the generated element is not located in the Weyl chamber we can simply discard it without comparing with the existing representatives and proceed to the next element.
In this sense we do not need to compare with the whole set of the Weyl group but instead only with the representatives.
After the simplifications are taken into account, finally we can generate the $E_7$ Weyl group within a few minutes.

\section*{Acknowledgements}

We are grateful to Yasuaki Hikida, Hiroshi Itoyama, Sung-Soo Kim, Kimyeong Lee, Marcos Marino, Shigeki Sugimoto, Yuji Sugimoto, Futoshi Yagi and Yasuhiko Yamada for valuable discussions and comments.
The work of S.M. is supported by JSPS Grant-in-Aid for Scientific Research (C) \#26400245.
S.M. would like to thank Yukawa Institute for Theoretical Physics at Kyoto University for warm hospitality.


\begin{thebibliography}{10}
\bibitem{ADKMV}
M.~Aganagic, R.~Dijkgraaf, A.~Klemm, M.~Marino and C.~Vafa,
``Topological strings and integrable hierarchies,''
Commun.\ Math.\ Phys.\  {\bf 261}, 451 (2006)
[hep-th/0312085].
%
\bibitem{ACDKV}
M.~Aganagic, M.~C.~N.~Cheng, R.~Dijkgraaf, D.~Krefl and C.~Vafa,
``Quantum Geometry of Refined Topological Strings,''
JHEP {\bf 1211}, 019 (2012)
[arXiv:1105.0630 [hep-th]].
%
\bibitem{NS}
N.~A.~Nekrasov and S.~L.~Shatashvili,
``Quantization of Integrable Systems and Four Dimensional Gauge Theories,''
arXiv:0908.4052 [hep-th].
%
\bibitem{MiMo}
A.~Mironov and A.~Morozov,
``Nekrasov Functions and Exact Bohr-Zommerfeld Integrals,''
JHEP {\bf 1004}, 040 (2010)
[arXiv:0910.5670 [hep-th]].
%
\bibitem{ABJM}
O.~Aharony, O.~Bergman, D.~L.~Jafferis and J.~Maldacena,
``N=6 superconformal Chern-Simons-matter theories, M2-branes and their gravity duals,''
JHEP {\bf 0810}, 091 (2008)
[arXiv:0806.1218 [hep-th]].
%
\bibitem{MP}
M.~Marino and P.~Putrov,
``ABJM theory as a Fermi gas,''
J.\ Stat.\ Mech.\  {\bf 1203}, P03001 (2012)
[arXiv:1110.4066 [hep-th]].
%
\bibitem{MPtop}
M.~Marino and P.~Putrov,
``Exact Results in ABJM Theory from Topological Strings,''
JHEP {\bf 1006}, 011 (2010)
[arXiv:0912.3074 [hep-th]].
%
\bibitem{DMP1}
N.~Drukker, M.~Marino and P.~Putrov,
``From weak to strong coupling in ABJM theory,''
Commun.\ Math.\ Phys.\  {\bf 306}, 511 (2011)
[arXiv:1007.3837 [hep-th]].
%
\bibitem{HKPT}
C.~P.~Herzog, I.~R.~Klebanov, S.~S.~Pufu and T.~Tesileanu,
``Multi-Matrix Models and Tri-Sasaki Einstein Spaces,''
Phys.\ Rev.\ D {\bf 83}, 046001 (2011)
[arXiv:1011.5487 [hep-th]].
%
\bibitem{DMP2}
N.~Drukker, M.~Marino and P.~Putrov,
``Nonperturbative aspects of ABJM theory,''
JHEP {\bf 1111}, 141 (2011)
[arXiv:1103.4844 [hep-th]].
%
\bibitem{FHM}
H.~Fuji, S.~Hirano and S.~Moriyama,
``Summing Up All Genus Free Energy of ABJM Matrix Model,''
JHEP {\bf 1108}, 001 (2011)
[arXiv:1106.4631 [hep-th]].
%
\bibitem{KEK}
M.~Hanada, M.~Honda, Y.~Honma, J.~Nishimura, S.~Shiba and Y.~Yoshida,
``Numerical studies of the ABJM theory for arbitrary N at arbitrary coupling constant,''
JHEP {\bf 1205}, 121 (2012)
[arXiv:1202.5300 [hep-th]].
%
\bibitem{HMO1}
Y.~Hatsuda, S.~Moriyama and K.~Okuyama,
``Exact Results on the ABJM Fermi Gas,''
JHEP {\bf 1210}, 020 (2012)
[arXiv:1207.4283 [hep-th]].
%
\bibitem{PY}
P.~Putrov and M.~Yamazaki,
``Exact ABJM Partition Function from TBA,''
Mod.\ Phys.\ Lett.\ A {\bf 27}, 1250200 (2012)
[arXiv:1207.5066 [hep-th]].
%
\bibitem{HMO2}
Y.~Hatsuda, S.~Moriyama and K.~Okuyama,
``Instanton Effects in ABJM Theory from Fermi Gas Approach,''
JHEP {\bf 1301}, 158 (2013)
[arXiv:1211.1251 [hep-th]].
%
\bibitem{CM}
F.~Calvo and M.~Marino,
``Membrane instantons from a semiclassical TBA,''
JHEP {\bf 1305}, 006 (2013)
[arXiv:1212.5118 [hep-th]].
%
\bibitem{HMO3}
Y.~Hatsuda, S.~Moriyama and K.~Okuyama,
``Instanton Bound States in ABJM Theory,''
JHEP {\bf 1305}, 054 (2013)
[arXiv:1301.5184 [hep-th]].
%
\bibitem{HMMO}
Y.~Hatsuda, M.~Marino, S.~Moriyama and K.~Okuyama,
``Non-perturbative effects and the refined topological string,''
JHEP {\bf 1409}, 168 (2014)
[arXiv:1306.1734 [hep-th]].
%
\bibitem{HLLLP2}
K.~Hosomichi, K.~M.~Lee, S.~Lee, S.~Lee and J.~Park,
``N=5,6 Superconformal Chern-Simons Theories and M2-branes on Orbifolds,''
JHEP {\bf 0809}, 002 (2008)
[arXiv:0806.4977 [hep-th]].
%
\bibitem{ABJ}
O.~Aharony, O.~Bergman and D.~L.~Jafferis,
``Fractional M2-branes,''
JHEP {\bf 0811}, 043 (2008)
[arXiv:0807.4924 [hep-th]].
%
\bibitem{AHS}
H.~Awata, S.~Hirano and M.~Shigemori,
``The Partition Function of ABJ Theory,''
PTEP {\bf 2013}, 053B04 (2013)
[arXiv:1212.2966 [hep-th]].
%
\bibitem{H}
M.~Honda,
``Direct derivation of "mirror" ABJ partition function,''
JHEP {\bf 1312}, 046 (2013)
[arXiv:1310.3126 [hep-th]].
%
\bibitem{MM}
S.~Matsumoto and S.~Moriyama,
``ABJ Fractional Brane from ABJM Wilson Loop,''
JHEP {\bf 1403}, 079 (2014)
[arXiv:1310.8051 [hep-th]].
%
\bibitem{HO}
M.~Honda and K.~Okuyama,
``Exact results on ABJ theory and the refined topological string,''
JHEP {\bf 1408}, 148 (2014)
[arXiv:1405.3653 [hep-th]].
%
\bibitem{HM}
M.~Honda and S.~Moriyama,
``Instanton Effects in Orbifold ABJM Theory,''
JHEP {\bf 1408}, 091 (2014)
[arXiv:1404.0676 [hep-th]].
%
\bibitem{MN1}
S.~Moriyama and T.~Nosaka,
``Partition Functions of Superconformal Chern-Simons Theories from Fermi Gas Approach,''
JHEP {\bf 1411}, 164 (2014)
[arXiv:1407.4268 [hep-th]].
%
\bibitem{MN2}
S.~Moriyama and T.~Nosaka,
``ABJM membrane instanton from a pole cancellation mechanism,''
Phys.\ Rev.\ D {\bf 92}, no. 2, 026003 (2015)
[arXiv:1410.4918 [hep-th]].
%
\bibitem{MN3}
S.~Moriyama and T.~Nosaka,
``Exact Instanton Expansion of Superconformal Chern-Simons Theories from Topological Strings,''
JHEP {\bf 1505}, 022 (2015)
[arXiv:1412.6243 [hep-th]].
%
\bibitem{HHO}
Y.~Hatsuda, M.~Honda and K.~Okuyama,
``Large N non-perturbative effects in $\mathcal{N}=4$ superconformal Chern-Simons theories,''
JHEP {\bf 1509}, 046 (2015)
[arXiv:1505.07120 [hep-th]].
%
\bibitem{MNN}
S.~Moriyama, S.~Nakayama and T.~Nosaka,
``Instanton Effects in Rank Deformed Superconformal Chern-Simons Theories from Topological Strings,''
JHEP {\bf 1708}, 003 (2017)
[arXiv:1704.04358 [hep-th]].
%
\bibitem{MNY}
S.~Moriyama, T.~Nosaka and K.~Yano,
``Superconformal Chern-Simons Theories from del Pezzo Geometries,''
JHEP {\bf 1711}, 089 (2017)
[arXiv:1707.02420 [hep-th]].
%
\bibitem{MS1}
S.~Moriyama and T.~Suyama,
``Instanton Effects in Orientifold ABJM Theory,''
JHEP {\bf 1603}, 034 (2016)
[arXiv:1511.01660 [hep-th]].
%
\bibitem{Hosp}
M.~Honda,
``Exact relations between M2-brane theories with and without Orientifolds,''
JHEP {\bf 1606} (2016) 123
[arXiv:1512.04335 [hep-th]].
%
\bibitem{Oosp}
K.~Okuyama,
``Orientifolding of the ABJ Fermi gas,''
JHEP {\bf 1603} (2016) 008
[arXiv:1601.03215 [hep-th]].
%
\bibitem{MS2}
S.~Moriyama and T.~Suyama,
``Orthosymplectic Chern-Simons Matrix Model and Chirality Projection,''
JHEP {\bf 1604} (2016) 132
[arXiv:1601.03846 [hep-th]].
%
\bibitem{MN5}
S.~Moriyama and T.~Nosaka,
``Orientifold ABJM Matrix Model: Chiral Projections and Worldsheet Instantons,''
JHEP {\bf 1606} (2016) 068
[arXiv:1603.00615 [hep-th]].
%
\bibitem{GHM1}
A.~Grassi, Y.~Hatsuda and M.~Marino,
``Topological Strings from Quantum Mechanics,''
Annales Henri Poincare {\bf 17}, no. 11, 3177 (2016)
[arXiv:1410.3382 [hep-th]].
%
\bibitem{OZ}
K.~Okuyama and S.~Zakany,
``TBA-like integral equations from quantized mirror curves,''
JHEP {\bf 1603}, 101 (2016)
[arXiv:1512.06904 [hep-th]].
%
\bibitem{KaMa}
M.~Marino and S.~Zakany,
``Matrix models from operators and topological strings,''
Annales Henri Poincare {\bf 17}, no. 5, 1075 (2016)
[arXiv:1502.02958 [hep-th]].
%
\bibitem{KMZ}
R.~Kashaev, M.~Marino and S.~Zakany,
``Matrix models from operators and topological strings, 2,''
Annales Henri Poincare {\bf 17}, no. 10, 2741 (2016)
[arXiv:1505.02243 [hep-th]].
%
\bibitem{KWY}
A.~Kapustin, B.~Willett and I.~Yaakov,
``Exact Results for Wilson Loops in Superconformal Chern-Simons Theories with Matter,''
JHEP {\bf 1003}, 089 (2010)
[arXiv:0909.4559 [hep-th]].
%
\bibitem{KuMo}
N.~Kubo and S.~Moriyama,
``Two-Point Functions in ABJM Matrix Model,''
JHEP {\bf 1805}, 181 (2018)
[arXiv:1803.07161 [hep-th]].
%
\bibitem{GAH}
D.~R.~Gulotta, J.~P.~Ang and C.~P.~Herzog,
``Matrix Models for Supersymmetric Chern-Simons Theories with an ADE Classification,''
JHEP {\bf 1201}, 132 (2012)
[arXiv:1111.1744 [hep-th]].
%
\bibitem{ADF}
B.~Assel, N.~Drukker and J.~Felix,
``Partition functions of 3d $\hat D$-quivers and their mirror duals from 1d free fermions,''
JHEP {\bf 1508}, 071 (2015)
[arXiv:1504.07636 [hep-th]].
%
\bibitem{MN4}
S.~Moriyama and T.~Nosaka,
``Superconformal Chern-Simons Partition Functions of Affine D-type Quiver from Fermi Gas,''
JHEP {\bf 1509}, 054 (2015)
[arXiv:1504.07710 [hep-th]].
%
\bibitem{IK4}
Y.~Imamura and K.~Kimura,
``N=4 Chern-Simons theories with auxiliary vector multiplets,''
JHEP {\bf 0810}, 040 (2008)
[arXiv:0807.2144 [hep-th]].
%
\bibitem{HKP}
M.~X.~Huang, A.~Klemm and M.~Poretschkin,
``Refined stable pair invariants for E-, M- and $[p, q]$-strings,''
JHEP {\bf 1311}, 112 (2013)
[arXiv:1308.0619 [hep-th]].
%
\bibitem{KNY}
K.~Kajiwara, M.~Noumi and Y.~Yamada,
``Geometric Aspects of Painlev\'e Equations,''
J.~Phys.~A: Math.~Theor. {\bf 50}, 073001 (2017)
[arxiv:1509.08186 [nlin]].
%
\bibitem{KMN2}
N.~Kubo, S.~Moriyama and T.~Nosaka, work in progress.
%
\bibitem{KimYagi}
S.~S.~Kim and F.~Yagi,
``5d E$_{n}$ Seiberg-Witten curve via toric-like diagram,''
JHEP {\bf 1506}, 082 (2015)
[arXiv:1411.7903 [hep-th]].
%
\bibitem{BBT}
F.~Benini, S.~Benvenuti and Y.~Tachikawa,
``Webs of five-branes and N=2 superconformal field theories,''
JHEP {\bf 0909}, 052 (2009)
[arXiv:0906.0359 [hep-th]].
%
\bibitem{CGM}
S.~Codesido, A.~Grassi and M.~Marino,
``Spectral Theory and Mirror Curves of Higher Genus,''
Annales Henri Poincare {\bf 18}, no. 2, 559 (2017)
[arXiv:1507.02096 [hep-th]].
%
\bibitem{BGT}
G.~Bonelli, A.~Grassi and A.~Tanzini,
``Quantum curves and $q$-deformed Painlev\'e equations,''
arXiv:1710.11603 {[}hep-th{]}.
%
\bibitem{GG}
A.~Grassi and J.~Gu,
``Argyres-Douglas theories, Painlev\'e II and quantum mechanics,''
arXiv:1803.02320 [hep-th].
%
\bibitem{IOY}
H.~Itoyama, T.~Oota and K.~Yano,
``Discrete Painleve system and the double scaling limit of the matrix model for irregular conformal block and gauge theory,''
arXiv:1805.05057 [hep-th].
%
\bibitem{AHK}
O.~Aharony, A.~Hanany and B.~Kol,
``Webs of (p,q) five-branes, five-dimensional field theories and grid diagrams,''
JHEP {\bf 9801}, 002 (1998)
[hep-th/9710116].
%
\bibitem{Mashimo}
K.~Mashimo,
``Weyl group of exceptional root system,'' Yuzawa 2013,
in japanese.
%
\end{thebibliography}
\end{document}